\documentclass[12pt]{article}

\usepackage{newtxtext,newtxmath}

\usepackage{graphicx}

\usepackage[letterpaper,margin=1in]{geometry}

\renewenvironment{abstract}
	{\quotation}
	{\endquotation}

\date{}

\makeatletter
\renewcommand{\fnum@figure}{\textbf{Figure \thefigure}}
\renewcommand{\fnum@table}{\textbf{Table \thetable}}
\makeatother

\usepackage{scicite}

\usepackage{url}

\newcommand{\araa}{Annu. Rev. Astron. Astrophys.} 
\newcommand{\aj}{Astron. J.} 
\newcommand{\apj}{Astrophys. J.} 
\newcommand{\apjl}{Astrophys. J. Lett.} 
\newcommand{\apjs}{Astrophys. J. Suppl. Ser.} 
\newcommand{\aap}{Astron. Astrophys.} 
\newcommand{\mnras}{Mon. Not. R. Astron. Soc.} 
\newcommand{\nat}{Nature} 
\newcommand{\physrep}{Phys. Rep.} 
\newcommand{\prl}{Phys. Rev. Lett.} 
\newcommand{\pasp}{Publ. Astron. Soc. Pac.} 
\newcommand{\ssr}{Space Sci. Rev.} 

\def\scititle{
	Harmonic brightness oscillations point to magnetar giant flares in rare short gamma-ray bursts
}
\title{\bfseries \boldmath \scititle}

\author{
    Run-Chao Chen$^{1,2\ast}$,
	Ore Gottlieb$^{3\ast}$,
    Bin-Bin Zhang$^{1,2\ast}$,\and
    Yuri Levin$^{4}$,
    Dong Lai$^{5,6}$,\and
    Jun Yang$^{7}$,
    Yi-Han Iris Yin$^{8,9}$,
    Suo-Ning Wang$^{1,2}$\and
	\small$^{1}$School of Astronomy and Space Science, Nanjing University, Nanjing 210093, China\and
	\small$^{2}$Key Laboratory of Modern Astronomy and Astrophysics (Nanjing University),\and \small Ministry of Education, China\and
	\small$^{3}$Department of Physics and Kavli Institute for Astrophysics and Space Research,\and 
    \small Massachusetts Institute of Technology, Cambridge, MA 02139, USA\and
	\small$^{4}$Department of Physics and Columbia Astrophysics Laboratory, Columbia University,\and
    \small Pupin Hall, New York, NY 10027, USA\and
	\small$^{5}$Tsung-Dao Lee Institute, Shanghai Jiao Tong University, Shanghai 201210, China\and
	\small$^{6}$Cornell Center for Astrophysics and Planetary Science, Department of Astronomy,\and
    \small Cornell University, Ithaca, NY 14853, USA\and
	\small$^{7}$Institute for Astrophysics, School of Physics, Zhengzhou University, Zhengzhou 450001, China\and
   \small$^{8}$Department of Physics, the University of Hong Kong, Pokfulam Road, Hong Kong, China\and
	\small$^{9}$The Hong Kong Institute for Astronomy and Astrophysics, The University of Hong Kong,\and
    \small Hong Kong, China\and
    \small$^\ast$Corresponding authors: R.-C. C. (\url{chrczxx@smail.nju.edu.cn}); O. G. (\url{ogot@mit.edu});\and 
    \small B.-B. Z. (\url{bbzhang@nju.edu.cn})\and
}

\begin{document} 

\maketitle
\begin{abstract} \bfseries \boldmath
Kilohertz quasi-periodic oscillations (QPOs) have been observed in a few rare gamma-ray bursts (GRBs), but their physical origin remains uncertain.
We reanalyze the two short GRBs with reported kilohertz QPOs, GRB~910711 and GRB~200415A, and find evidence for possible harmonic-like organization around a characteristic frequency scale near 360~Hz.
The similar harmonic-like pattern in GRB~910711 and in the confirmed magnetar giant flare (MGF) event GRB~200415A, including features extending above 4~kHz, points to timing phenomenology on frequency scales relevant to magnetar oscillation modes. The temporal and spectral similarities between the two events and the difficulty of explaining these features in merger-driven scenarios favor an MGF-like interpretation of GRB~910711. Such timing phenomenology may therefore provide an additional observational handle, complementary to spectral, temporal, and localization information, for distinguishing rare MGF-like transients from merger-driven GRBs.
\end{abstract}

\section*{Introduction}
\noindent Brightness oscillations are widely observed during thermonuclear X-ray bursts from accreting neutron stars\cite{1996ApJ...469L...9S,2000ARA&A..38..717V,2008ApJS..179..360G,2012ARA&A..50..609W}. More recently, oscillations with broadly similar characteristics have been reported in a few short GRBs\cite{2021Natur.600..621C,2023Natur.613..253C,2026ApJ...998..289C}. Unlike those in X-ray bursts, short-GRB oscillations are exceptionally rare\cite{2023Natur.613..253C,2026ApJ...998..289C}, typically lasting only tens of milliseconds and appearing at frequencies of a few kilohertz\cite{2021Natur.600..621C,2023Natur.613..253C}. Given that short GRBs may arise from two physically distinct channels, merger-driven GRBs\cite{2014ARA&A..52...43B} and magnetar giant flares (MGFs)\cite{2006ApJ...640..849N}, the origin of their oscillations should be considered within both physical scenarios. In merger-driven scenarios, kilohertz oscillations are often discussed in the context of central-engine dynamics\cite{2019ApJ...884L..16C,2023Natur.613..253C,2025NatAs...9.1701C,2026ApJ...998..289C}, whereas in MGFs they may be linked to neutron-star oscillations or magnetospheric variability\cite{2007MNRAS.377..159L,2011MNRAS.418..659L,2021Natur.600..621C}. Nevertheless, the similarity in duration and spectral hardness between some extragalactic MGFs and merger-driven GRBs\cite{2005MNRAS.362L...8L} renders the two populations difficult to distinguish without precise localizations\cite{2005Natur.434.1098H,2005Natur.434.1107P}. In the absence of decisive progenitor evidence, establishing the origin of high-frequency QPOs therefore requires considering physically distinct origins across both channels.

\subsection*{The Interpretive Tension in GRB~910711}

Among short GRBs with reported kilohertz brightness oscillations, GRB~910711 represents a particularly compelling case. It exhibits rare sub-millisecond temporal structures\cite{1992Natur.359..217B} and remains the strongest candidate for kilohertz QPOs among 724 short GRBs analyzed under a Bayesian model-comparison framework, where the signal was interpreted in the context of post-merger hypermassive neutron-star (HMNS) dynamics\cite{2023Natur.613..253C}. Importantly, the high signal-to-noise ratio of the burst enables spectral measurements with the Spectroscopy Detectors (SDs) of the Burst and Transient Source Experiment (BATSE)\cite{1992NASCP3137...26F} onboard the \textit{Compton Gamma Ray Observatory} (\textit{CGRO}). Using SDs with favorable incident geometry (see Materials and Methods), we find that the time-integrated spectrum of GRB~910711 is comparably described by both blackbody and cutoff power-law models under Bayesian information criterion (BIC) comparison (Table~\ref{tab1})\cite{1978AnSta...6..461S}.

This spectral behavior introduces a tension in the interpretation of the reported kilohertz oscillations. Although the spectrum is comparably well fitted by a cutoff power-law model often used to describe merger-driven GRB prompt emission, the inferred low-energy photon index is unusually hard ($\alpha\approx0.87$). Such a value substantially exceeds the expectations of conventional optically thin nonthermal emission models and lies at the extreme end of the BATSE GRB spectral distribution\cite{1998ApJ...506L..23P,2013ApJS..208...21G,2015PhR...561....1K}. Moreover, the SD light curves closely trace the main burst emission (Fig.~\ref{efig1}\textbf{A}), making unresolved spectral evolution a less likely explanation for the hard time-integrated spectrum. A natural alternative is that the observed radiation originates from an optically thick photosphere\cite{2017SSRv..207...87B,2015PhR...561....1K}, in which case the hard spectrum is less problematic. However, this interpretation introduces a separate difficulty: the observed radiation would predominantly emerge from scattering-rich photospheric regions far from the central engine\cite{2017SSRv..207...87B,2015PhR...561....1K}. Preserving coherent kilohertz modulation from the proposed HMNS engine through such a region would therefore require an additional mechanism that either maintains the modulation across the photosphere or regenerates it near the emission site.

Motivated by this tension, we explore whether an alternative MGF interpretation can provide a physically plausible context for the reported oscillations in GRB~910711. We therefore compare it with GRB~200415A, a burst identified as an extragalactic MGF from NGC~253\cite{2021Natur.589..207R,2021Natur.589..211S}. GRB~200415A provides a natural benchmark because it exhibits a remarkably similar phenomenology, including reported very-high-frequency brightness oscillations\cite{2021Natur.600..621C}, unusually hard spectral behavior, and exceptionally rapid variability extending to sub-millisecond timescales\cite{2021Natur.589..207R}.

\section*{Results}
\subsection*{Comparison between GRB~910711 and GRB~200415A}

For GRB~910711, the timing analysis is performed using data from the Large Area Detectors (LADs) of BATSE (see Materials and Methods). For GRB~200415A, we draw on observations from the Burst Alert Telescope (BAT)\cite{2005SSRv..120..143B} on the \textit{Neil Gehrels Swift Observatory} and the Atmosphere--Space Interactions Monitor (ASIM)\cite{2019SSRv..215...23O} on the \textit{International Space Station}.

\subsubsection*{Temporal structure similarities}

Fig.~\ref{fig1} compares the light curves of the two bursts. When aligning the 0.1-ms binned light curves by their peak times (an alignment strategy applied consistently throughout all subsequent time-coordinate adjustments) (see Materials and Methods) and segmenting the time series using a Bayesian block algorithm\cite{2013ApJ...764..167S}, both show a clear morphological resemblance: a bright, sharp initial spike lasting $\approx$10~ms, followed by a fainter tail of $\approx$100--200~ms. Sub-millisecond structures are evident during the spike's rising edge. In GRB~910711, the shortest resolved feature has a half-duration of $\approx$90~$\mu$s, while in GRB~200415A, features as short as $\approx$65~$\mu$s are detected. With a more permissive prior, a component as brief as $\approx$26~$\mu$s emerges in GRB~200415A (Fig.~\ref{fig1}\textbf{C}).

To demonstrate that GRB~910711 represents a fainter analog of GRB~200415A, we perform temporal profile fitting of both events (see Materials and Methods). Modeling each burst as comprising two fast-rise and exponential-decay components (FRED; representing the initial spike and decaying tail), we find that the BAT light curve of GRB~200415A, when scaled to a luminosity distance of $\approx$8.75~Mpc, provides a fainter analogue of GRB~910711, retaining the characteristic short spike followed by a weaker and shorter-lived extended tail (Fig.~\ref{efig2}, Table~\ref{etab1}). This resemblance is consistent with the interpretation of GRB~910711 as an extragalactic MGF serendipitously detected by BATSE.

\subsubsection*{Low-frequency excesses associated with the reported kilohertz brightness oscillations}

Beyond their similar light-curve morphology and rapid spike structures, both GRB~910711 and GRB~200415A show kilohertz brightness oscillations during the initial spike\cite{2021Natur.600..621C,2023Natur.613..253C}. To compare these oscillation features, we use the aligned light curves shown in Fig.~\ref{fig1}\textbf{C} and compute Leahy-normalized power spectra\cite{1983ApJ...266..160L} for the [96, 106]~ms interval, alongside Lomb--Scargle periodograms evaluated on an oversampled frequency grid\cite{2018ApJS..236...16V}. As presented in Fig.~\ref{fig2}, GRB~910711 and GRB~200415A show prominent low-frequency excesses at 360 and 350~Hz, respectively, while GRB~910711 additionally exhibits a high-frequency excess at 2100~Hz, close to the $\approx$2132~Hz oscillation reported in GRB~200415A\cite{2021Natur.600..621C}. Given the $\approx$10~ms duration of the spike, the frequency difference between these excesses lies below the characteristic Fourier resolution scale.

Because the low-frequency excesses are identified in very short, rapidly variable spikes and resolved only through an oversampled frequency grid, we first test whether they could arise from pulse-envelope variability and finite-duration leakage. The light curves within [92, 112]~ms were fitted with multiple FRED pulse decompositions plus background to capture the rapidly varying spike envelopes, with model complexity selected using BIC (see Materials and Methods). This procedure yields a two-FRED description for GRB~910711 and a three-FRED description for GRB~200415A (Fig.~\ref{efig3}, Table~\ref{etab2}). From posterior samples of these envelope models, $10^5$ synthetic light curves were generated and subjected to the same analysis procedure, with the low-frequency search performed within $\pm100$~Hz of the observed peak frequency (see Materials and Methods). As shown in Fig.~\ref{fig2}, under this envelope-only null hypothesis, the observed low-frequency excesses are recovered with probabilities of $p\approx0.051$ for GRB~910711 and $p\approx0.001$ for GRB~200415A, indicating that they are not readily produced by pulse-envelope variability alone.

We next examine the robustness of the previously unreported $\approx$2100~Hz high-frequency excess in GRB~910711 using weighted wavelet Z-transform (WWZ)\cite{1996AJ....112.1709F} spectrograms. As shown in Fig.~\ref{efig4}, $\approx$2160~Hz high-frequency excesses are present in both bursts, remain visible across instruments and energy bands, and are consistently aligned with the initial spikes over a broad frequency range tested. To quantify the significance of these high-frequency excesses, the power spectra were modeled using Lorentzian components to represent additional QPOs above a red-noise plus white-noise continuum (see Materials and Methods). Relative to the continuum-only null, adding two Lorentzian components substantially improves the description of the observed power spectra, favoring a two-QPO interpretation (likelihood-ratio tests yield $\mathrm{TS}=13.4$ for GRB~910711 and $\mathrm{TS}=33.9$ for GRB~200415A). Consistently, the posterior distributions of the preferred models recover high-frequency excesses near $\approx$2160~Hz in both bursts (Table~\ref{etab3}).

Interestingly, we also generate $10^5$ synthetic light curves from the QPO-informed power-spectrum models using the Timmer--Koenig method\cite{1995A&A...300..707T} and repeat the same low-frequency search (see Materials and Methods). The apparent significances of the low-frequency excesses decrease substantially, to $p\approx0.155$ for GRB~910711 and $p\approx0.334$ for GRB~200415A (Fig.~\ref{fig2}). This result suggests that, while the low-frequency excesses are not readily reproduced by pulse-envelope variability alone, they become less unexpected once the reported kilohertz brightness oscillations are included, motivating further investigation of a possible connection between the low- and high-frequency excesses.

\subsubsection*{Suggestive evidence for a 360-Hz harmonic-like organization}

Although the low-frequency excesses have reduced significance under the QPO-informed null model, another suggestive feature is that both the low- and high-frequency excesses cluster near integer multiples of a characteristic frequency scale of $\approx$360~Hz (Table~\ref{etab3}). Motivated by this numerical correspondence, we analyze 20 frequency bands defined by successive integer multiples of the characteristic frequency scale near 360~Hz in the initial spikes of both bursts using WWZ analysis (see Materials and Methods). As shown in Fig.~\ref{fig3}, multiple high-frequency excesses near integer multiples of $\approx$360~Hz are directly observed in the WWZ maps and remain temporally aligned with the initial spikes. This behavior suggests a possible connection among these high-frequency excesses, potentially consistent with nonlinear coupling, and motivates further investigation using higher-order spectral analysis.

At the most stringent level, we test whether these harmonics exhibit nonlinear coupling. The relationships among the ``candidate'' harmonic nodes become clearer when examined using the bispectrum, which quantifies nonlinear coupling between frequencies\cite{2013MNRAS.435.3547M}. In both bursts, the bispectral magnitude is concentrated at frequency pairs that correspond to the harmonics observed in the spectrograms (Fig.~\ref{fig3}). For instance, in GRB~200415A the $12\times360 = 4320$~Hz feature is consistent with nonlinear coupling between two $6\times360 = 2160$~Hz components, which themselves originate from nonlinear coupling between the $2\times360=720$~Hz and $4\times360=1440$~Hz high-frequency excesses. This hierarchy suggests that, when the low-frequency excess identified in the LSP analysis is treated as a characteristic frequency scale near 360~Hz, the observed high-frequency excesses in both bursts become naturally connected through approximate harmonic relations.

In addition, Fig.~\ref{fig3} reveals a naturally testable harmonic-like structure, in which the observed high-frequency excesses approximately follow $n\times f_0$ with $f_0\approx360\pm100$~Hz while extending to harmonics as high as $n=12$. Because small differences in the characteristic frequency scale are progressively amplified at higher harmonics, this structure enables a more stringent test of frequency preference beyond the nominal Fourier resolution associated with the 10~ms spike duration. Using the harmonic nodes identified in the spectrograms, we evaluate the preferred characteristic frequency within the same $360\pm100$~Hz interval as an internal cross-check of the harmonic-like structure (see Materials and Methods). Despite differences in the specific harmonic nodes recovered in the two bursts, both favor a characteristic frequency close to 360~Hz (Fig.~\ref{efig5}\textbf{A\&D}).

Still adopting the FRED-envelope null model introduced above, we further examine the robustness of the harmonic-like structure using $10^5$ synthetic light curves (see Materials and Methods). First, the preferred characteristic frequency inferred from the full harmonic-node structure shows no strong preference for 360~Hz in the simulations, suggesting that the observed characteristic frequency scale near 360~Hz is unlikely to arise solely from short-window effects and the variable envelopes (Fig.~\ref{efig5}\textbf{B\&E}). We then compare the maximum bispectral coupling strength recovered within the same $360\pm100$~Hz interval in each realization with the observed couplings at $\approx$360~Hz, finding probabilities of $p\approx1.3\times10^{-4}$ for GRB~910711 and $p\approx4.5\times10^{-4}$ for GRB~200415A (Fig.~\ref{efig5}\textbf{B\&E}). Finally, the individual harmonic nodes provide non-independent consistency checks that support their inclusion within the observed harmonic-like structure (Fig.~\ref{efig5}\textbf{C\&F}).

\subsection*{Distance constraints and MGF plausibility of GRB~910711}

Using the spectral parameters inferred (Table~\ref{tab1}), we place GRB~910711 in the empirical $E_{\rm p,z}$--$E_{\rm iso}$ plane over an assumed redshift range spanning nearby galaxies to cosmological GRBs ($z\approx0.0005$--10). As shown in Fig.~\ref{efig6}, if interpreted as a merger-driven GRB, GRB~910711 appears comparatively underluminous for its spectral peak energy relative to merger-driven GRBs. By contrast, when compared with known MGFs, including previously proposed extragalactic candidates\cite{2020ApJ...903L..32Z}, its spectral energetics remain consistent with an MGF-like population out to luminosity distances of $\approx$55~Mpc.

Motivated by this empirical upper distance scale, we systematically search for nearby galaxies within the BATSE localization region using the GLADE+ catalog\cite{2022MNRAS.514.1403D}. Within the allowed distance range, 17 dwarf galaxies are identified as potential hosts (Table~\ref{etab4}). In the GLADE+ catalog, the photometric properties of all candidate galaxies are classified as being more consistent with ongoing star formation than with passive or quenched stellar populations, broadly consistent with environments expected to host young magnetars capable of producing giant flares\cite{2021ApJ...907L..28B}.

\section*{Discussion}
\subsection*{Physical implications}

Similar to GRB~200415A, the unusually hard spectrum of GRB~910711 is broadly compatible with an MGF framework, in which the prompt hard X-ray/$\gamma$-ray spike is commonly interpreted as radiation emerging from a hot pair--photon fireball and magnetospheric pair plasma formed in an ultra-strong magnetic field\cite{2001ApJ...561..980T}. In such a picture, quasi-thermal or Comptonized spectral shapes may arise naturally and are broadly compatible with the spectral properties inferred for GRB~910711.

The harmonic-like timing organization reported here may likewise admit a magnetar-related interpretation, as high-frequency QPO-like signals observed in Galactic giant flares have often been discussed in the context of seismic or magnetospheric oscillations triggered during flare activity\cite{2005ApJ...628L..53I,2005ApJ...632L.111S,2006ApJ...637L.117W,2006ApJ...653..593S}. However, such signals have primarily been detected in the decaying tails of Galactic giant flares rather than during the prompt spike. The excitation of the observed high-frequency nodes and, more importantly, their coupling to magnetospheric radiation processes during the initial spike, therefore, remain uncertain and require further theoretical investigation.

At the same time, the characteristic frequency scale near 360~Hz identified in both bursts provides a phenomenological framework that connects the previously reported oscillations extending above 4~kHz to a lower characteristic frequency scale. This scale is broadly comparable to frequencies of high-frequency QPO-like signals reported in Galactic giant-flare tails, including features near $\approx$625~Hz\cite{2005ApJ...634L.153P,2014ApJ...793..129H}. Although the harmonic nodes excited in the two events differ, and although the Galactic signals were detected in flare tails rather than prompt spikes, this frequency-scale similarity may help contextualize the extreme high-frequency oscillations observed during the initial spikes of extragalactic MGFs. The physically relevant oscillation scale could also correspond to a lower common divisor of the observed 360~Hz, a possibility that awaits further theoretical clarification.

Taken together, these results do not uniquely establish an MGF origin for GRB~910711, but they make it the more economical interpretation of the available evidence. A merger-driven interpretation must simultaneously account for the unusually hard spectrum, the preservation or regeneration of coherent kilohertz modulation through a scattering-rich photospheric emission region, the close phenomenological resemblance to the confirmed MGF GRB~200415A, and the shared harmonic-like timing organization. By contrast, in an MGF framework these properties arise within a single physical context involving a magnetar-powered prompt spike, magnetospheric pair plasma, and neutron-star or magnetospheric oscillations.

\subsection*{Observational implications}

Kilohertz brightness oscillations have been reported in rare short high-energy transients, interpreted within both merger-driven GRBs and MGFs, but their physical origin remains uncertain\cite{2021Natur.600..621C,2023Natur.613..253C,2025NatAs...9.1701C,2026ApJ...998..289C}. GRB~910711 highlights the ambiguity that can arise when precise localization or secure host identification is unavailable. Nevertheless, the combination of its unusually hard spectrum, rapid temporal structure, close resemblance to the confirmed MGF GRB~200415A, distance-scale plausibility, and shared harmonic-like timing organization makes an MGF-like interpretation more natural than a merger-driven one, even though it is not uniquely established. More broadly, if similar harmonic-like timing organization is identified in future short GRBs, systematic assessment of its compatibility with different physical-origin scenarios, together with spectral, temporal, and localization information, may provide an additional observational handle for distinguishing rare MGF-like transients from merger-driven GRBs using prompt high-energy data alone. Such information could complement localization constraints while guiding rapid follow-up strategies, since nearby MGFs and merger-driven GRBs are expected to motivate substantially different observational priorities.

\section*{Materials and Methods}

\subsection*{Instrument and data}
\noindent\textbf{BATSE.} The BATSE onboard the \textit{CGRO} comprised eight LADs and eight SDs, providing nearly all-sky coverage for high-energy transients. Each LAD had an effective area of 2025~cm$^2$ and a nominal energy range of 20~keV–1.9~MeV. LADs were optimized for timing and, during transients, operated in Time-Tagged Event (TTE) mode, recording individual photon arrival times with 2~$\mu$s resolution in four broad energy channels ($\approx$25–60, 60–110, 110–325, and $>$325~keV). In comparison, SDs had a smaller effective area (126~cm$^2$) but a broader energy coverage (10~keV to $>$100~MeV, depending on gain) and finer energy resolution, delivering spectra with up to 256 channels for detailed spectroscopy\cite{BATSE_url}.

\noindent\textbf{Swift-BAT.} BAT\cite{2005SSRv..120..143B} is a wide-field coded-aperture instrument with a CdZnTe detector plane of total sensitive area $\approx$5240~cm$^2$, sensitive over 15--350~keV and capable of recording photons up to $\approx$500~keV. BAT normally operates in a real-time triggering mode: onboard rate and image algorithms identify significant count-rate increases, localize the burst, and can command rapid follow-up observations. However, high-time-resolution event data (TTE, with 100~$\mu$s resolution) are only telemetered around such triggers. The Gamma-ray Urgent Archiver for Novel Opportunities (GUANO)\cite{2020ApJ...900...35T} is a commanding pipeline that enables targeted recovery of stored BAT event data in response to external alerts. This system allows sensitive searches for sub-threshold or untriggered bursts by retrieving event-mode data around the epoch of interest, even when no onboard trigger occurred.

\noindent\textbf{ASIM-MXGS.} The Modular X- and Gamma-Ray Sensor (MXGS)\cite{2019SSRv..215...23O} is the high-energy instrument of the ASIM payload aboard the \textit{International Space Station}. MXGS was designed to study terrestrial gamma-ray flashes but is also sensitive to astrophysical transients, including MGFs. It comprises two detector subsystems: the Low-Energy Detector (LED; 1024~cm$^2$, 20–400~keV, 1~$\mu$s time resolution) and the High-Energy Detector (HED; 900~cm$^2$, 200–40,000~keV, 28~ns time resolution), providing high-time-resolution measurements across a broad energy range.

\noindent\textbf{GRB~910711.} GRB~910711 was detected exclusively by BATSE at 09:31:52~UTC on 11 July 1991 (trigger 512) and was initially reported as the shortest GRB detected by the mission\cite{1992NASCP3137...26F}. The refined localization places the source at RA~$=217.5^{\circ}$, Dec~$=-16.8^{\circ}$ (J2000), with an estimated positional uncertainty of about 6$^{\circ}$\cite{1992Natur.359..217B}. Five of the eight LADs were triggered; however, subsequent analyses showed that LAD3 and LAD7 predominantly recorded Earth-scattered photons owing to unfavorable viewing geometries, introducing significant propagation delays\cite{1992Natur.359..217B}. To minimize such contamination and preserve intrinsic timing, we therefore excluded LAD3 and LAD7 from the temporal analysis.

For temporal analysis, we used TTE data from LAD4, LAD5, and LAD6, the detectors with the most favorable source angles and minimal contamination from atmospheric scattering. These data were retrieved from the BATSE TTE \texttt{ASCII} archive. For spectral analysis, we used Spectroscopy Time-Tagged Event (STTE) data from the BATSE trigger archive. In the STTE mode used here, spectra are provided in 252 energy channels.

\noindent\textbf{GRB~200415A.} GRB~200415A triggered the \textit{Fermi} Gamma-ray Burst Monitor (GBM)\cite{2009ApJ...702..791M} and the \textit{Fermi} Large Area Telescope (LAT)\cite{2009ApJ...697.1071A} at 08:48:05~UT on 15 April 2020\cite{2020GCN.27579....1F,2020GCN.27586....1O}. Owing to its exceptional brightness, the burst was also detected by multiple high-energy instruments, including Swift-BAT (with TTE data recovered via GUANO), ASIM-MXGS\cite{2020GCN.27622....1M}, Konus-Wind\cite{2020GCN.27596....1F}, and the Imager on Board the INTEGRAL Satellite (IBIS)\cite{2020GCN.27627....1P}. An enhanced localization from the Inter-Planetary Network of satellites places the burst at RA = $11.88^{\circ}$ and Dec = $-25.263^{\circ}$ (J2000), with a positional uncertainty within a 17 acrmin$^2$ region\cite{2021Natur.589..207R}. The burst was spatially associated with the nearby galaxy NGC~253\cite{2005MNRAS.361..330R} and classified as an extragalactic MGF\cite{2021Natur.589..207R,2021Natur.589..211S}. However, due to its extreme brightness, much of the GBM data suffered saturation and data loss, making timing analysis unreliable\cite{2021Natur.589..207R}. Therefore, in this study, we used only publicly available data from Swift-BAT\cite{2021Natur.589..207R} and ASIM\cite{2021Natur.600..621C}.

As GRB~200415A lay outside BAT's coded field of view and would not have triggered the instrument autonomously\cite{2021Natur.589..207R}, we used GUANO\cite{2020ApJ...900...35T} to retrieve full-resolution TTE data. All TTE records for GRB~200415A from the LED and HED of ASIM are publicly available as reported in ref.\cite{2021Natur.600..621C}. For temporal analysis, we rebinned the event data into fixed-width time bins within the energy ranges appropriate for each instrument, assigning each bin a timestamp at its midpoint. Unlike Swift-BAT data, which include \texttt{FITS} files specifying event energies, the ASIM data are provided in raw counts per detector channel\cite{2021Natur.600..621C} and were converted to physical energies in keV using a quadratic calibration function derived from pre-launch measurements\cite{2019SSRv..215...23O}. Given that the temporal resolution limit of Swift-BAT is $\approx$100~$\mu$s, we used BAT data for temporal analysis requiring resolution above this limit, and ASIM data for analyses requiring finer temporal resolution or energy coverage above 400~keV. For spectral analysis, we adopted the results reported in previous studies\cite{2020ApJ...899..106Y,2020ApJ...903L..32Z,2021Natur.589..207R}.

\subsection*{Temporal analysis}
\noindent\textbf{Light curves comparison.} To facilitate a comparison between the observed light curves of GRB~910711 and GRB~200415A, we apply the following procedures:
\begin{enumerate}
\item We restrict the energy range to 60--325~keV, corresponding to channels 2 and 3 of the BATSE-LADs, and overlaps with the sensitivity ranges of the other gamma-ray instruments used in this study.
\item The reference times for GRB~910711 and GRB~200415A from different instruments are shifted such that the peak of each light curve, calculated using 100-$\mu$s binning (the minimum bin size of Swift-BAT), is aligned to that of GRB~910711, which is set to 0.10145~s.
\end{enumerate}
In this study, the reference time of GRB~200415A was adjusted to align with that of GRB~910711 by applying the above procedure at each step. The only variation was in the choice of energy ranges. The procedure and the event timestamps of GRB~910711 were not modified, remaining consistent with the \texttt{ASCII} file of the BATSE TTE data.

\begin{figure}
\centering
\includegraphics[width=0.75\textwidth]{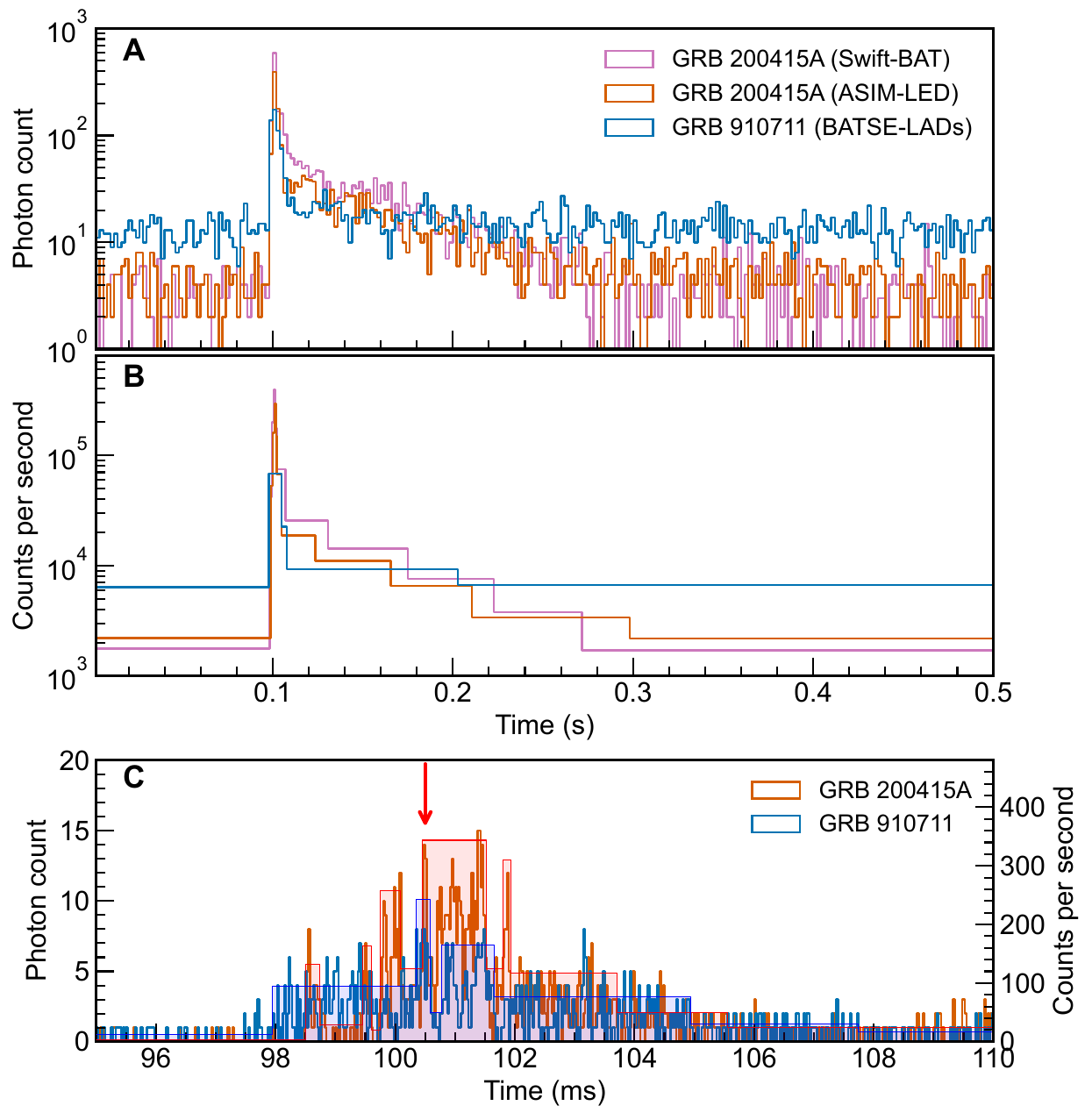} 
\caption{\textbf{Temporal similarities between GRB~910711 and GRB~200415A.}
Time axes are aligned to the peak of each burst, as identified in 0.1-ms resolution data. 
(\textbf{A}) Multi-instrument light curves in the 60--325~keV band, shown at 2-ms resolution. 
(\textbf{B}) Bayesian block segmentation of the light curves in panel~A, highlighting the temporal structures of each burst. The prior probability\cite{2013ApJ...764..167S} for the Bayesian block algorithm was set to $p_0 = 0.001$ to avoid over-segmentation of the global light curve morphology. Both bursts show a sharp, millisecond-scale initial spike followed by a weaker, sub-second tail. 
(\textbf{C}) Initial spikes of GRB~910711 (BATSE-LADs, 20~keV--1.9~MeV) and GRB~200415A (ASIM-LED, 20--400~keV) at 25~$\mu$s resolution. Bayesian blocks under the prior $p_0 = 0.1$ are overlaid as shaded regions: blue for GRB~910711, red for GRB~200415A. Both bursts show similar sub-millisecond variability with multiple pulses. The finest structure in GRB~200415A is revealed only when the prior is relaxed to 0.2, marked by the red arrow.}
\label{fig1}
\end{figure}

\begin{figure}
\centering
\includegraphics[width=\textwidth]{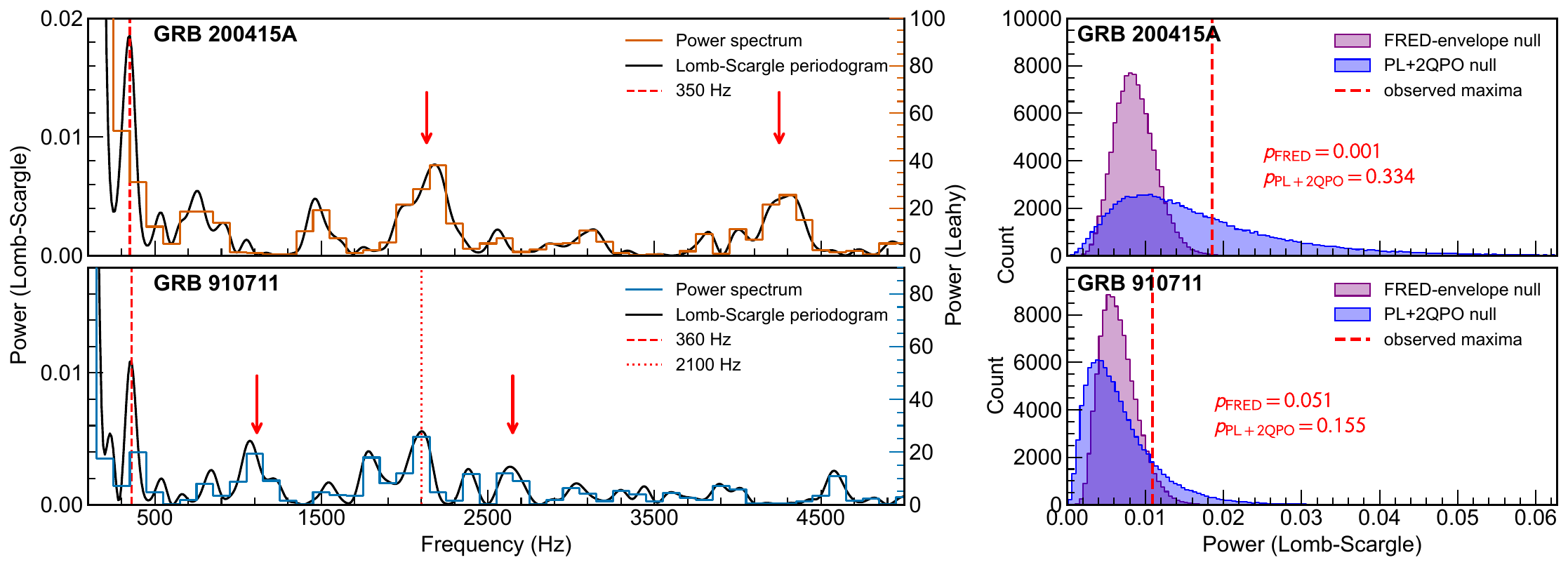} 
\caption{\textbf{Low-frequency excesses associated with the reported kilohertz brightness oscillations in GRB~910711 and GRB~200415A.}
\textbf{Left:} Leahy-normalized power spectra\cite{1983ApJ...266..160L} and Lomb--Scargle periodograms computed from 2-$\mu$s binned light curves in the interval [96, 106]~ms (see Fig.~\ref{fig1}\textbf{C}), shown for GRB~200415A (top) and GRB~910711 (bottom). The displayed frequency range is 100--5000~Hz, and the Lomb--Scargle periodograms are evaluated on an oversampled frequency grid with 10-Hz spacing. Both bursts exhibit prominent low-frequency excesses near 350 and 360~Hz (red dashed lines). Red arrows indicate previously reported kilohertz brightness oscillations\cite{2021Natur.600..621C,2023Natur.613..253C}; GRB~910711 additionally shows an excess near 2100~Hz, close to the $\approx$2132~Hz oscillation reported in GRB~200415A.
\textbf{Right:} Monte Carlo assessment of the low-frequency excesses. Histograms show the null distributions of the maximum Lomb--Scargle power recovered within the same $260$--$460$~Hz search interval from $10^5$ synthetic light curves generated under two assumptions: a FRED-envelope null model (purple), representing pulse-envelope variability alone, and a PL+2QPO null model (blue), incorporating the previously reported kilohertz oscillatory excesses in the power spectrum. The red dashed line marks the observed maximum Lomb--Scargle power. Under the FRED-envelope null, the low-frequency excesses are not readily reproduced, with probabilities of $p_{\rm FRED}\approx0.001$ for GRB~200415A and $0.051$ for GRB~910711. Once the previously reported kilohertz excesses are incorporated, the corresponding probabilities increase to $p_{\rm PL+2QPO}\approx0.334$ and $0.155$, respectively, suggesting that the low-frequency excesses become less unexpected in the presence of the reported high-frequency oscillatory structure.}
\label{fig2}
\end{figure}

\begin{figure}
\centering
\includegraphics[width=0.65\textwidth]{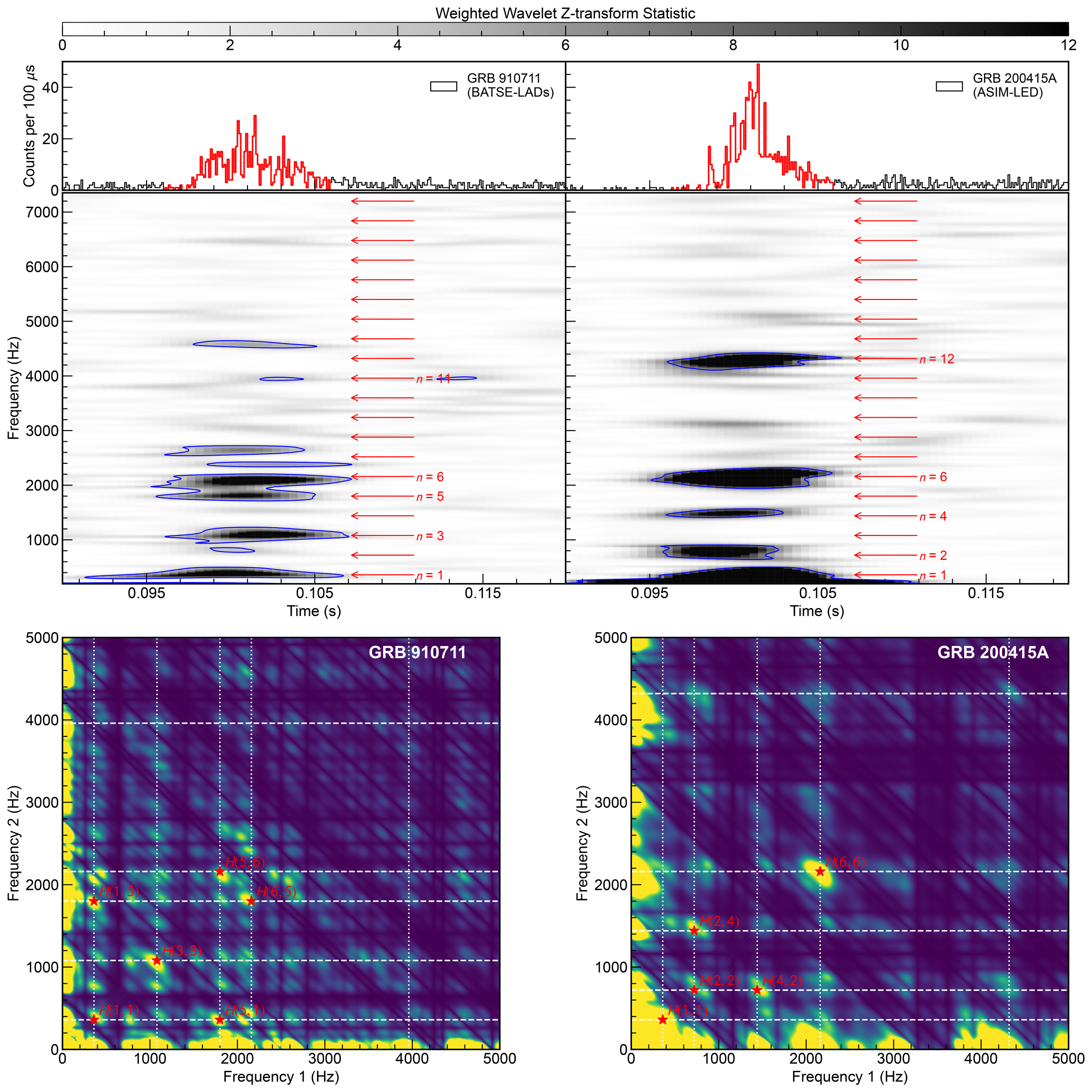} 
\caption{\textbf{Harmonic analysis of the shared 360-Hz oscillation feature in GRB~910711 and GRB~200415A.}
\textbf{Top:} Light curves binned at 0.1~ms without energy selection are shown in the top panels, with corresponding spectrograms from 2-$\mu$s-binned data below. Blue contours mark the top 95\% of the test statistics. Harmonic candidates appear during the initial spikes at integer multiples $n\times f$ with $f=360$~Hz simultaneously, with $n=1, 3, 5, 6, 11$ for GRB~910711 and $n=1, 2, 4, 6, 12$ for GRB~200415A. 
\textbf{Bottom:} Bispectrum of the initial spike of GRB~910711 and GRB~200415A over the interval [96, 106]~ms (red-shaded region in top panel), computed with 50-$\mu$s bins. The frequency grid is refined to 10~Hz resolution through zero padding\cite{1986nras.book.....P}, and a Hamming window\cite{1978IEEEP..66...51H} is applied to reduce red-noise leakage. Brighter regions in the bispectrum indicate larger bispectral magnitudes, corresponding to stronger nonlinear coupling. Symbols $H(x,y)$ (equivalent to $H(y,x)$) denote harmonic features at the frequency pair $(x\times f,\, y\times f)$, representing nonlinear coupling between two oscillations. Such coupling generates power at $x\times f$, $y\times f$, and their sum $(x+y)\times f$ in the Fourier spectrum. Frequency pairs closely associated with the harmonic peaks in spectrograms are marked with red stars, indicating the presence of a consistent harmonic system in both events.}
\label{fig3}
\end{figure}

Fig.~\ref{fig1} compares the light curves of GRB~910711 and GRB~200415A, with temporal structures extracted using the Bayesian block algorithm\cite{2013ApJ...764..167S}. In this method, the prior parameter $p_0$ sets the false-positive probability for detecting a change point in a purely random Poisson process: smaller values of $p_0$ result in fewer blocks, highlighting only the most robust structures, while larger values allow finer subdivisions, revealing more detailed features but increasing the chance of spurious detections. Accordingly, we adopted $p_0 = 0.001$ to analyze the overall light curve and capture the primary structure of each burst (Fig.~\ref{fig1}\textbf{B}), ensuring that only statistically significant blocks are identified. To resolve the fine sub-millisecond structures within the initial spikes, we increased $p_0$ to 0.1--0.2, allowing the algorithm to detect rapid variations that would otherwise be merged at lower prior thresholds (Fig.~\ref{fig1}\textbf{C}).

As shown by the Bayesian blocks, even with a very small prior of $p_0 = 0.001$, a weak tail lasting approximately 100~ms is detected above the background level (Fig.~\ref{fig1}\textbf{B}). The overall light-curve morphology of GRB~910711 and GRB~200415A in a common energy range is thus similar, each consisting of a bright initial spike lasting about 10~ms, followed by a weaker decaying tail of 100–200~ms. Using a more permissive prior of $p_0 = 0.1$ applied to the full energy range of the BATSE-LADs and ASIM-LED, we determined the minimum variability timescale (MVT), defined here as the half-duration of the shortest block, and found that it falls well below a millisecond for both bursts. Although $p_0 = 0.1$ does not even resolve the fastest components during the brightest part of GRB~200415A (Fig.~\ref{fig1}\textbf{C}), the MVTs of both bursts are nearly two orders of magnitude smaller than typical values in the MVT distribution of short GRBs\cite{2015ApJ...811...93G,2023A&A...671A.112C}. This extreme variability in the initial spikes further suggests a possible origin distinct from the standard merger scenario.

\noindent\textbf{Model fitting of the light curves.} Based on the strong morphological resemblance between the light curves of GRB~910711 and GRB~200415A, we hypothesize that both bursts may originate from a similar physical scenario. GRB~910711 could be interpreted as a fainter analogue of GRB~200415A, serendipitously detected by BATSE owing to its higher sensitivity. In the following, we perform model fitting on the light curves of both events to quantitatively test this hypothesis.

Fig.~\ref{efig2}\textbf{A} shows the 60--325~keV light curve of GRB~200415A from Swift-BAT, binned at 200~$\mu$s. To describe the temporal profile, we adopt a model consisting of two fast-rise exponential-decay (FRED) components. A single-FRED model is insufficient: fitting with only one FRED leaves a systematic residual excess around 0.104~s in both the Swift-BAT and ASIM data. The excess is inconsistent with Poisson fluctuations and suggests an additional structure in the rising phase (0.102--0.105~s). Adding a second FRED component removes the excess and yields a more accurate description of the observed pulse morphology. A single FRED model is defined as:
\begin{equation}
F(t) =
\left\{
\begin{array}{ll}
A \left( \frac{t}{t_{\text{peak}}} \right)^{t_{\text{peak}}/\tau_{\text{rise}}}, & \text{if } t < t_{\text{peak}} \\
A \exp\left( -\frac{(t - t_{\text{peak}})}{\tau_{\text{decay}}} \right), & \text{if } t \geq t_{\text{peak}}
\end{array}
\right.
\label{eq_fred}
\end{equation}
where $F(t)$ denotes the observed photon counts as a function of time $t$, $A$ is the amplitude, $t_{\text{peak}}$ is the peak time, $\tau_{\text{rise}}$ and $\tau_{\text{decay}}$ are the characteristic rise and decay timescales, respectively.

To fit the observed light curve of GRB~200415A within [0, 0.5]~s, we perform Markov Chain Monte Carlo simulations (MCMCs) using the \texttt{Python} package \texttt{emcee}\cite{2010CAMCS...5...65G, 2013PASP..125..306F}. The fitted model consists of double FRED components plus a constant background level ($C$). A Gaussian likelihood is assumed for the double FRED signal, while a Poisson likelihood is used for the background component. The fitting employs 32 walkers, each generating 5,000 samples, with the first 20\% treated as burn-in and discarded.

Table~\ref{etab1} summarizes the fit results. As the posterior distributions of all parameters converge well, we report the median values as the best-fit results, with 1$\sigma$ uncertainties defined by the 16th and 84th percentiles. The best-fit model is shown in Fig.~\ref{efig2}\textbf{A}, along with simulated light curves generated by adding Poisson noise to the model prediction. As illustrated in the insets, the double-FRED-plus-constant model, denoted as $F_1 + F_2 + C$, effectively reproduces the overall temporal features of the observed light curve of GRB~200415A.

In the second step, we assume that GRB~910711 is a fainter analog of GRB~200415A. Under this assumption, we use the best-fit model $F_1 + F_2 + C$ as a template and scale its amplitudes to reproduce the light curve of GRB~910711. This process is controlled by three key parameters: The amplification factor $a$ scales the net signal intensity of GRB~200415A (i.e., the double FRED component $F_1 + F_2$) and reflects three contributing factors: the intrinsic luminosity difference arising from the burst's physical origin, the observed luminosity variation due to distance, and the instrumental sensitivity. The instrumental background factor $b$, which modifies the observed background level of GRB~200415A (i.e., the constant $C$). Assuming GRB~200415A occurs at a different distance, this term reflects only instrumental sensitivity. The intrinsic timescale modifier $c$, which uniformly scales both the rise and decay timescales of the burst. This parameter is governed solely by the underlying physical mechanism and is independent of both distance and instrumental sensitivity.

With these considerations, the best-fit model of GRB~200415A, originally expressed as $F_1 + F_2 + C$, is modified to: 
\begin{equation}
a \times \left(F_1\{c\} + F_2\{c\} \right) + b \times C,
\end{equation}
where $F\{c\}$ denotes a FRED profile whose timescales are scaled as $\tau_{\text{rise}}/c$ and $\tau_{\text{decay}}/c$.
We note that parameters $a$ and $b$ are not fully independent, as instrumental sensitivity affects both the observed signal and background. We choose to separate them to allow for a more balanced weight allocation during fitting. The parameter $c$ is admittedly a simplified treatment, especially given the poorly understood physical mechanism of MGFs. We apply the same $c$ to both FRED components to avoid introducing additional physical bias into this mathematical test.

Although omitting the parameter $c$ might seem preferable for model simplicity, we retain it to account for non-negligible intrinsic temporal variations among MGFs. For instance, both GRB~200415A and GRB~231115A originated from approximately the same distance (3.5~Mpc), yet GRB~231115A exhibits significantly lower intensity and notably shallower rise and decay slopes than GRB~200415A\cite{2024ApJ...963L..10Y}. This suggests that intrinsic physical differences, beyond distance or instrumental effects, may influence both the observed flux (associated with factor $a$) and the temporal evolution timescale (associated with factor $c$). However, in this study, we do not further investigate the underlying mechanisms and neglect any potential correlation between $a$ and $c$.

To perform the model fitting, we adopt the best-fit parameters of GRB~200415A, treating only $a$, $b$, and $c$ as free parameters when fitting the observed light curve of GRB~910711 within the interval $[0, 0.5]$~s. For this step, we reduce the number of ensemble walkers from 32 to 16 while keeping the sample size unchanged in the MCMC sampling. The resulting fit parameters are also summarized in Table~\ref{etab1}.

As shown in Fig.~\ref{efig2}\textbf{B}, the light curve of GRB~910711 is consistent with a rescaled version of the GRB~200415A template under our assumption. Bayesian block analysis of the simulated light curve also reveals a two-phase temporal structure (short spike + extended tail), matching the observed morphology of GRB~910711. Based on our assumption, the change in background level is attributed to differences in instrumental sensitivity between Swift-BAT and BATSE-LADs, the best-fit value $b = 1.419$ implies a $\approx40\%$ enhancement for BATSE. Under this circumstance, if the best-fit rescaling parameter $a = 0.227$ solely reflects distance effects (ignoring intrinsic physical differences), the inferred ratio suggests GRB~910711 was $6.25$ times fainter than GRB~200415A. Thus, based on the quantitative model fitting, if GRB~200415A were placed at a luminosity distance of $8.75$~Mpc, its light curve would match the observed shape of GRB~910711. 

It should be emphasized that the purpose of this exercise is primarily phenomenological: to test whether the overall temporal morphology of GRB~910711 can be approximately reproduced by a rescaled version of GRB~200415A under a simplified MGF-motivated assumption. The model is not intended to represent the detailed physical processes governing MGFs, nor should it be interpreted as a quantitative estimator of the luminosity distance or intrinsic energetics of GRB~910711. Instead, the inferred scaling serves only as an illustrative indication that the global light-curve structures of the two events are broadly compatible under a common phenomenological picture.

\subsection*{Spectral fitting of GRB~910711}
We performed the spectral fitting of GRB~910711 using the STTE data from BATSE's Spectroscopy Detectors (SDs). Among the available detectors, SD4, SD5, and SD6 recorded the burst, with angular offsets of $30^{\circ}$, $52^{\circ}$, and $55^{\circ}$ from the source direction, and $159^{\circ}$, $135^{\circ}$, and $78^{\circ}$ from the Earth, respectively. Because the relatively small Earth angle of SD6 increases the likelihood of atmospheric scattering, we carried out spectral fits both including and excluding SD6 to assess the impact of detector selection on the results.

For each detector, we used the Gamma-ray Data Tools (GDT)\cite{GDT-CORE, GDT-CGRO} to extract total and background spectra over the interval [0.09,\,0.11]~s, with the timing referenced to the LADs \texttt{ASCII} event data (Fig.~\ref{efig1}\textbf{A}), together with the corresponding response matrices. Spectral fitting was performed with \texttt{bayspec}\cite{Yang2022Natur, Yang2023ApJ, Yin2025ApJ,bayspec_url}, employing the PGSTAT statistic\cite{Arnaud1996ASPC} to evaluate the likelihood between model and data within the effective energy range of 25--1800~keV. The models tested included a blackbody, a single power law, a cutoff power law, a band function, and a multi-color blackbody\cite{Hou2018ApJ,2020ApJ...899..106Y}.

We first applied these models to the combined SD4+SD5 dataset and to the SD4+SD5+SD6 dataset. In all cases, the photon count spectrum from SD6 showed significant deviations from those of SD4 and SD5, and from the corresponding best-fit models, particularly at lower energies. As an illustration, Fig.~\ref{efig1}\textbf{B\&C} show the blackbody fits to SD4+SD5 and SD4+SD5+SD6, respectively. These results indicate that the SD6 spectrum is substantially contaminated by atmospheric scattering. Consequently, in the following analysis we restrict spectral fitting to SD4 and SD5. 

The fitting results and associated statistical metrics are given in Table~\ref{tab1}. Model comparison based on the BIC\cite{1978AnSta...6..461S} indicates that the blackbody model provides the best description, with a temperature of $kT = 216^{+43}_{-37}~{\rm keV}$. The cutoff power law model is the second-best fit, characterized by a very hard low-energy photon index of $\alpha = 0.87^{+0.67}_{-0.68}$.

\subsection*{Quasi-periodic oscillation analysis}
\noindent\textbf{Periodogram analysis.} The light curves of GRB~910711 and GRB~200415A were sampled at uniform intervals, allowing standard fast-Fourier-transform (FFT) analysis. However, the short duration of the analyzed segments ($\Delta T \approx10$~ms) and the steep red-noise continuum produce strong low-frequency power that contaminates neighboring frequencies through spectral leakage. In an FFT, the Rayleigh resolution is $\Delta f_{\rm R}=1/\Delta T$, and the use of a rectangular window of length $\Delta T$ spreads low-frequency power into higher bins on scales comparable to $\Delta f_{\rm R}$. This leakage, combined with the large red-noise amplitude, extends the influence of low-frequency variance into the hundreds of hertz, potentially obscuring narrow high-frequency signals (Fig.~\ref{fig2}). Alternative approaches such as window functions\cite{1978IEEEP..66...51H} or zero-padding\cite{1986nras.book.....P} can mitigate leakage or improve visual interpolation, but none of these approaches improves the true resolving power.

By contrast, the Lomb--Scargle periodogram\cite{2018ApJS..236...16V} evaluates sinusoidal fits at arbitrary trial frequencies and, with controlled oversampling, can localize narrow features that would otherwise be spread across neighboring FFT bins. We computed the Lomb--Scargle periodogram with the \texttt{Python} package \texttt{astropy.timeseries.LombScargle}\cite{2022ApJ...935..167A} using its default ``standard'' normalization (power interpreted as the fractional reduction in variance relative to a constant model), and evaluated it on a grid oversampled by a factor of ten relative to the Rayleigh resolution ($\Delta f=\Delta f_{\rm R}/10$). As shown in Fig.~\ref{fig2}, this approach reveals a narrow peak around 360~Hz in both bursts that is absent or strongly suppressed in the FFT owing to red-noise leakage and grid discretization.
We emphasize that such oversampling does not improve the intrinsic frequency resolution or by itself establish statistical significance, and may under strong red noise produce enhanced local maxima owing to interpolation of correlated power. For this reason, all subsequent significance estimates were evaluated against synthetic surrogate light curves analyzed using the identical oversampled Lomb–Scargle procedure, thereby explicitly incorporating any bias introduced by oversampling, spectral leakage, and red-noise contamination into the null distribution.

\noindent\textbf{Spectrogram analysis.} To investigate whether the $\approx$360-Hz feature persists over time and whether it connects to harmonic structures, we applied the Weighted Wavelet Z-transform (WWZ)\cite{1996AJ....112.1709F} for time--frequency analysis. Standard approaches such as the short-time Fourier transform (STFT)\cite{Grochenig2001} apply a fixed window and are constrained by the time--frequency trade-off, which is suboptimal for millisecond bursts with rapid, short-lived oscillations. The WWZ instead uses Gaussian weighting centered on each trial time to fit localized sinusoids, making it well suited to detecting narrow, high-frequency modulations embedded in strongly variable light curves.

Formally, the light curve is projected onto a localized sinusoidal basis
\begin{equation}
f(t) = A\cos[2\pi f (t-\tau)] + B\sin[2\pi f (t-\tau)] + C,
\end{equation}
where $A$, $B$, and $C$ are regression coefficients, $f$ is the trial frequency, and $\tau$ is the temporal centre. Each point is weighted by a Gaussian kernel,
\begin{equation}
w_i = \exp[-\alpha (t_i-\tau)^2],
\end{equation}
with $\alpha = c_w (2\pi f)^2$, where $c_w$ is a tunable constant that controls the effective temporal resolution. The resulting WWZ statistic is
\begin{equation}
\mathrm{WWZ} = \frac{(N_{\rm eff}-3)V_y}{2(V_x-V_y)},
\end{equation}
with $V_x$ the weighted variance of the data, $V_y$ the residual variance after fitting, and $N_{\rm eff}$ the effective number of weighted points\cite{1996AJ....111..541F}. Larger values correspond to stronger localized periodicity.

Alongside WWZ, we also computed the weighted wavelet amplitude (WWA),
\begin{equation}
\mathrm{WWA}(f,\tau) = \sqrt{A^2+B^2},
\end{equation}
which directly quantifies the oscillation amplitude at each $(f,\tau)$. While WWZ measures the significance of a localized periodicity relative to background variance, WWA highlights the absolute modulation strength. Together, these diagnostics provide complementary views of the time--frequency behavior of the bursts.

For the analysis, we concentrated on harmonics of the candidate $f_0=360$~Hz fundamental revealed in the periodogram analysis. Central frequencies were set at $n\times f_0$ ($n=1,2,\dots,21$), with boundaries at the geometric means of adjacent harmonics. Light curves were binned at $2~\mu$s across the full BATSE-LAD and ASIM-LED energy ranges, and spectrograms computed for [0.09, 0.12]~s, covering the initial spike where the 360-Hz feature is strongest. The Gaussian decay constant for each frequency was set as $c_w = 1/(\Delta T \times n \times f_0)^2$, and the temporal step in $\tau$ was 0.5~ms. Calculations were carried out with the \texttt{Python} package \texttt{libwwz} (\url{https://github.com/ISLA-UH/libwwz}), within each harmonic interval, the transform was evaluated on a frequency resolution of 10~Hz, and the resulting time--frequency spectrograms were concatenated to form the final map. The resulting WWZ spectrograms (Fig.~\ref{fig3}) reveal activity at integer multiples of 360~Hz in both bursts. In GRB~910711, peaks appear simultaneously at $n\times360$~Hz with $n=1,3,5,6,11$. In GRB~200415A, the fundamental overlaps with elevated red-noise continuum power and is therefore not cleanly separable, but peaks at $n\times360$~Hz with $n=2,4,6,12$ remain prominent. The components at $3\times360$~Hz, $6\times360$~Hz, and $12\times360$~Hz are consistent with earlier reports\cite{2021Natur.600..621C,2023Natur.613..253C}, and notably, a common peak at $6\times360$~Hz is shared by both bursts, consistent with the additional peak near 2100~Hz seen in the Lomb--Scargle periodogram (Fig.~\ref{fig2}).

To test the consistency of the $\approx$2160-Hz ($6\times360$~Hz) feature across instruments and energy ranges, we performed a uniform high-frequency analysis over 500--4900~Hz with 10-Hz resolution, using 100-$\mu$s binned light curves from BATSE, Swift-BAT, and ASIM. The 2160-Hz harmonic is clearly recovered at both low energies ($\lesssim350$~keV) and high energies ($\gtrsim300$~keV), demonstrating that the signal persists across instruments and energy bands (Fig.~\ref{efig4}). 
Although the $\approx$2160~Hz component was not emphasized in previous studies\cite{2023Natur.613..253C}, this is expected given the different analysis objectives. Earlier work searched for the most significant kilohertz-scale QPO candidates with a freely varying centroid frequency and, in GRB~910711, identified a highly significant signal near $\approx$2649~Hz in the BATSE high-energy channels. Under such a framework, emphasizing the strongest localized excess is natural, and additional features at nearby frequencies need not be highlighted; this is consistent with Fig.~\ref{efig4}\textbf{E}. By contrast, our motivation here is to examine whether a harmonic-like organization emerges around the candidate $\approx$360~Hz scale. In this context, a component near $\approx$2160~Hz ($6\times360$~Hz) is consistently recovered across instruments, energy ranges, and time--frequency analyses, suggesting that it may represent a recurrent element of the candidate harmonic system. Because the previously reported $\approx$2649~Hz feature appears strongly energy selective and occupies a distinct frequency scale (Fig.~\ref{efig4}\textbf{E}), we do not attempt an extended interpretation of its physical relation to the candidate harmonic structure here.

To conclude from the spectrogram analysis, the simultaneous appearance of consistent peaks suggests a possible harmonic-like structure associated with the candidate $\approx$360-Hz scale revealed in the periodogram analysis. However, temporal coincidence alone cannot establish whether these candidate harmonics arise from genuine nonlinear coupling. A higher-order Fourier analysis is therefore required to test whether the observed frequency components exhibit coupling behavior consistent with a shared harmonic system in both bursts.

\noindent\textbf{Bispectrum analysis.} Building on the spectrogram results, which indicated simultaneous peaks at integer multiples of $\approx$360~Hz, we require a diagnostic that can test whether these components are genuinely coupled rather than coincident fluctuations. Standard periodogram or time--frequency analyses measure variance as a function of frequency but do not capture phase relations between components. The bispectrum\cite{2013MNRAS.435.3547M} provides this capability by testing whether Fourier components at $(f_1,f_2)$ coherently combine to produce power at $f_1{+}f_2$.

Formally, for a real, uniformly sampled time series $x(t)$ with Fourier transform $X(f)$, the (ensemble) bispectrum is
\begin{equation}
B(f_1,f_2)\;\equiv\;\mathbb{E}
\left[
X(f_1)\,X(f_2)\,X^{*}(f_1{+}f_2)
\right],
\label{eq:bispec_def}
\end{equation}
where $\mathbb{E}[\cdot]$ denotes an ensemble average and ${}^{*}$ the complex conjugate. In the present short-duration application, meaningful ensemble averaging is not feasible, and we therefore estimate the bispectrum directly from the full burst interval as
\begin{equation}
\hat B(f_1,f_2)
=
X(f_1)\,X(f_2)\,X^{*}(f_1{+}f_2).
\label{eq:bispec_est}
\end{equation}
For real-valued light curves we restrict the analysis to the non-redundant domain $f_1 \ge 0$, $f_2 \ge 0$, and $f_1{+}f_2 \le f_{\rm Nyq}=1/(2\Delta t)$ (the Nyquist frequency). Throughout, we report the bispectral magnitude $|\hat B(f_1,f_2)|$, i.e.\ the modulus of the complex bispectrum. This quantity reflects the strength of quadratic coupling between frequency components, independent of the associated phase relation.

To test whether the simultaneous components in the spectrograms form a genuine harmonic system, we focus on $|\hat B|$, which quantifies the strength of quadratic coupling between frequencies. Complementary diagnostics include the bispectral phase, which tests whether coupled components are phase-locked, and the bispectral coherence, which provides a unit-normalized measure of coupling strength. In principle, all three are needed for a complete characterization of nonlinear interactions. In this study, however, the spikes last only $\approx$10~ms, yielding very short time series with limited oscillation cycles at the relevant frequencies.
In long stationary signals, ensemble averaging is commonly estimated using multiple statistically independent segments. However, under the short-duration conditions considered here, meaningful subdivision into many independent segments is not feasible.
Under such conditions, the corresponding phase- and coherence-based estimators have low signal-to-noise and are highly sensitive to spectral leakage from the steep red-noise background, making phase- and coherence-based estimators comparatively noisy and difficult to interpret. The magnitude, by contrast, depends only on the modulus of the coupled terms and is therefore less affected by these limitations. We thus restrict our analysis to $|\hat B|$, which provides a conservative but reliable indicator of nonlinear coupling in these bursts.

We utilized the \texttt{stingray} package\cite{2019ApJ...881...39H} and analyzed light curves from [0.096,\,0.106]~s with 50-$\mu$s bins, considering the full energy ranges of BATSE and ASIM-LED. To improve frequency sampling, we applied zero-padding\cite{1986nras.book.....P}, which increases visual resolution by a factor of ten without altering the intrinsic Rayleigh resolution, and used a Hamming window\cite{1978IEEEP..66...51H} to suppress spectral leakage from strong low-frequency power. The Hamming taper is particularly advantageous here, as it reduces sidelobes that would otherwise contaminate higher-frequency couplings in the presence of steep red noise. The resulting bispectral maps in Fig.~\ref{fig3} reveal distinct couplings consistent with the spectrogram features: for GRB~910711, enhanced bispectral features appear at the frequency pairs $(f,5f)$, $(3f,3f)$, and $(5f,6f)$; for GRB~200415A, enhanced bispectral features appear at $(2f,2f)$, $(2f,4f)$, and $(6f,6f)$, with $f=360$~Hz. All of these frequency pairs coincide with the harmonic components seen in the WWZ spectrograms during the same time window, supporting an interpretation in which these components are consistent with nonlinear coupling and may together participate in a common harmonic-like structure.

\subsection*{Synthetic light-curve generation and statistical tests.}

To assess whether the candidate $\approx$360-Hz harmonic organization could arise from deterministic burst morphology, stochastic variability, and counting noise, we generated synthetic light curves that reproduce the observed temporal structure and noise properties. Because the analyzed spikes are short, strongly non-stationary transients with asymmetric pulse profiles, we first modelled the burst envelopes using phenomenological FRED components (Eq.~\ref{eq_fred}) to capture the deterministic light-curve morphology. We then characterized the residual stochastic variability in the Fourier domain using power-spectral models, comparing a red-noise continuum with an alternative model including additional narrow oscillatory components. These ingredients were subsequently combined with Poisson sampling to construct surrogate light curves for statistical evaluation of the periodogram, spectrogram, and bispectrum analyses.

\noindent\textbf{Light-curve synthesis from burst envelope fitting.} To address the possibility that deterministic but non-periodic pulse-envelope structures in a strongly non-stationary $\approx$10-ms transient could themselves generate narrow spectral features, we first constructed an envelope-only null model for each burst. In this null model, the prompt light curves of GRB~910711 and GRB~200415A were described using phenomenological superpositions of FRED components, and synthetic light curves were generated by applying Poisson fluctuations to the fitted envelopes.

For GRB~910711, we used the full-energy-band BATSE light curve constructed from LAD4+LAD5, while for GRB~200415A we adopted the full-energy-band ASIM-LED light curve. Both light curves were analyzed with a temporal resolution of 50~$\mu$s, which remains smaller than the minimum variability timescale identified from the Bayesian-block analysis (Fig.~\ref{fig1}~\textbf{C}) while avoiding excessive temporal refinement that would reduce the contrast between intrinsic variability and counting noise. For each burst, we considered models consisting of one to four FRED components plus a constant background term and performed posterior sampling using MCMCs with \texttt{emcee}, adopting 500 walkers and 1200 sampling steps, while discarding the first 200 steps as burn-in.

To determine the preferred envelope complexity, an initial fitting stage was performed in which all FRED peak times were allowed to vary within a common interval of 96--106~ms. This intentionally permissive setup was designed to allow different components to compete for the same temporal structures under potentially overfitted conditions, thereby enabling empirical model comparison through the BIC. The rise and decay timescales were restricted to values larger than 20~$\mu$s, ensuring that the minimum allowed variability scale remained smaller than half of the shortest Bayesian-block segment. Since the amplitudes of the burst substructures differ only moderately, the FRED amplitudes were modeled in linear rather than logarithmic space. The prior settings for the amplitude, rise timescale, and decay timescale of each FRED component were the same as those listed in Table~\ref{etab2}.

For GRB~910711, the resulting BIC values for the one- to four-component models are $-304.9$, $-334.2$, $-320.1$, and $-296.9$, respectively, favoring a two-component envelope model. For GRB~200415A, the corresponding BIC values are $-1620.7$, $-1644.9$, $-1660.3$, and $-1661.1$. Although the four-component model formally yields the lowest BIC value, the improvement relative to the three-component model is negligible ($\Delta{\rm BIC}=0.8 < 2$), indicating no meaningful statistical preference for an additional component. We therefore adopt the simpler three-component description following the principle of parsimony. Based on these comparisons, a 2-FRED model for GRB~910711 and a 3-FRED model for GRB~200415A were adopted in the subsequent analysis.

A second fitting stage was then performed using the preferred model for each burst, with non-overlapping peak-time intervals imposed to reduce parameter competition and obtain more stable parameter estimates. The resulting best-fit parameters are summarized in Table~\ref{etab2}.

To generate synthetic burst envelopes, we randomly sampled the top 20\% of posterior realizations from the preferred models and generated $10^5$ synthetic light curves by applying Poisson fluctuations to each realization\cite{2013MNRAS.433..907E}. As shown in Fig.~\ref{efig3}~\textbf{A}, the observed burst envelopes can be effectively reproduced across the posterior parameter space. For synthetic light curves generated at time resolutions finer than 50~$\mu$s, the amplitudes were rescaled proportionally to preserve the expected photon counts per time bin. It should be noted that, these phenomenological envelope realizations are not intended as physical descriptions of the burst emission, but rather as a conservative null construction to assess the level of stochastic variability that can arise from sharply peaked, strongly non-stationary burst morphology in the absence of periodic modulation.

\noindent\textbf{Light-curve synthesis from power spectra.} While the phenomenological burst-envelope fits described above reproduce the strongly non-stationary morphology of the spikes, they do not constrain the stochastic variability expected in the Fourier domain. To assess whether the harmonic organization anchored to the $\approx$360~Hz characteristic scale can arise from red-noise fluctuations alone, we further modeled the short-timescale variability using power-spectrum descriptions and generated synthetic light curves consistent with the inferred stochastic properties.

We considered two phenomenological models for the power spectrum: a red-noise continuum model,
\begin{equation}
P_{\rm PL+C}(f)=A f^{-\alpha}+C,
\end{equation}
and an alternative model including two Lorentzian excesses,
\begin{equation}
P_{\rm PL+C+2QPO}(f)
=
A f^{-\alpha}+C
+
\sum_{i=1}^{2}
\frac{A_i (w_i/2)^2}
{(f-f_i)^2+(w_i/2)^2},
\end{equation}
Here, $A$ is the continuum normalization, $\alpha$ is the red-noise spectral index, and $C$ is the constant white-noise component. The Lorentzian terms describe previously reported kilohertz-scale QPO-like excesses, with amplitudes $A_i$, centroid frequencies $f_i$, and widths $w_i$.

The motivation for introducing the two-QPO terms is not to establish independent QPO detections, which have already been reported in previous studies\cite{2021Natur.600..621C,2023Natur.613..253C}, but rather to reassess their consistency in the specific context of the candidate harmonic organization investigated here. In particular, because the harmonic interpretation developed in this work predicts additional structure near integer multiples of $\approx$360~Hz, reproducing the stochastic background alone is insufficient for constructing realistic null simulations. Instead, we test whether a stochastic model incorporating the previously reported excesses provides a more self-consistent description of the observed variability and therefore a more appropriate basis for evaluating the robustness of the harmonic system.

Both bursts were analyzed using $2~\mu$s-binned light curves, corresponding to the highest available temporal resolution of BATSE and therefore providing the longest Fourier baseline and most stable characterization of the short-timescale stochastic variability. Before periodogram construction, the mean count rate was removed such that the zero-frequency (DC) component was excluded. Model fitting employed the Whittle likelihood\cite{2022ApJS..259...32H},
\begin{equation}
\ln\mathcal{L}
=
-\sum_i
\left[
\ln P_{\rm model}(f_i)
+
\frac{P_{\rm obs}(f_i)}
{P_{\rm model}(f_i)}
\right],
\end{equation}
sampled using MCMCs with \texttt{emcee}. For the PL+C model, we adopted 500 walkers and 1200 sampling steps while discarding the first 200 steps as burn-in. For the PL+C+2QPO model, we used 800 walkers and 875 sampling steps and discarded the first 250 steps.

The adopted priors and posterior constraints are summarized in Table~\ref{etab3}. For GRB~910711, because the $\approx$2100~Hz feature was not explicitly reported in previous analyses\cite{2023Natur.613..253C}, its centroid frequency was allowed to vary broadly within 1500--3000~Hz and was allowed to converge freely. For all other Lorentzian components, centroid-frequency priors were adopted from previously reported QPO frequencies together with uncertainty propagation based on the reported widths\cite{2021Natur.600..621C,2023Natur.613..253C}. The Lorentzian widths were constrained to remain larger than 50~Hz, motivated by the 10~ms duration of the burst spikes, while also preventing excessively broad continuum-like features by maintaining approximately resolved oscillatory structures with quality factor $Q\equiv f/w\gtrsim2$.

For both bursts, the PL+C+2QPO model yielded higher maximum Whittle likelihoods than the PL+C model (Table~\ref{etab3}), corresponding to likelihood-ratio improvements
\begin{equation}
TS
=
2
\left(
\ln\mathcal{L}_{\rm PL+C+2QPO}
-
\ln\mathcal{L}_{\rm PL+C}
\right)
\end{equation}
of $TS=13.4$ for GRB~910711 and $TS=33.9$ for GRB~200415A. Consistent with previous reports\cite{2021Natur.600..621C,2023Natur.613..253C}, these results favor a description including localized QPO-like excesses in addition to a red-noise continuum and motivate their inclusion when constructing stochastic surrogate light curves.

Based on the posterior distributions of the fitted models, synthetic light-curve generation proceeded as follows:
\begin{enumerate}
\item Based on the posterior distributions of the PL+C and PL+C+2QPO models, we retained the top 20\% of samples ranked by posterior probability, thereby selecting the highest-posterior realizations for stochastic simulations. For each burst and each model, this procedure yielded $10^5$ posterior realizations, which were subsequently used to generate synthetic light curves.
\item For each retained parameter set, synthetic light curves were generated using the Timmer--Koenig algorithm\cite{1995A&A...300..707T} (hereafter T--K synthesis). In the Fourier domain, random complex amplitudes were drawn from a complex Gaussian distribution with variance prescribed by the target power spectrum model, and Hermitian symmetry was imposed such that the inverse Fourier transform yielded real-valued stochastic time series. The resulting realizations reproduce the variability prescribed by the fitted power-spectrum model, including the corresponding red- and white-noise components and, where applicable, localized QPO-like excesses.
\item The continuous stochastic series $x$ was converted into photon-counting data through Poisson sampling\cite{2013MNRAS.433..907E}. A constant offset $b$ was applied such that the mean of $\max(x+b,\varepsilon)$ matched the observed mean count rate $\mu_{\rm obs}$, where $\varepsilon$ is a small positive floor. Negative fluctuations were clipped only at this stage, yielding expected counts per bin $\lambda=\max(x+b,\varepsilon)$, from which discrete counts were drawn according to $y\sim{\rm Poisson}(\lambda)$.
\end{enumerate}
Because T--K realizations are Gaussian stochastic processes, occasional negative excursions are unavoidable and clipping introduces a small distortion of the high-frequency power spectrum. Across our simulations, clipping remained moderate but model dependent. For the pure PL+C model, a substantial fraction of posterior realizations could exhibit clipping levels exceeding $\approx$35\%, whereas the inclusion of QPO-like components generally reduced clipping, with typical realizations clustering around $\approx$10--15\%. We therefore regard clipping as a secondary systematic effect that primarily impacts the highest frequencies.

To evaluate how well the stochastic surrogates reproduce the observed variability, we compared the simulated and observed power spectra. As shown in Fig.~\ref{efig3}~\textbf{B--D}, simulations based on the PL+C model tend to overestimate the low-frequency variance while reproducing the observed localized excesses poorly. By contrast, the PL+C+2QPO model provides a substantially improved description of both the broad red-noise continuum and the previously reported kilohertz-scale excesses.

\noindent\textbf{Statistical tests for the Lomb--Scargle periodogram.}
To assess whether the low-frequency excesses identified in the Lomb--Scargle periodograms could arise from stochastic variability and counting statistics alone, we performed Monte Carlo simulations using the synthetic light curves described above. For each burst and each stochastic model (pulse-envelope-only and QPO-informed), $10^5$ surrogate light curves were generated and analyzed using the same Lomb--Scargle procedure adopted for the observed data, including identical oversampling (factor ten).

Because the low-frequency excesses are identified through an oversampled frequency grid and are associated with very short ($\approx$10~ms) spike intervals, their apparent prominence may be influenced by finite-duration leakage and local red-noise fluctuations. To account for these effects in a manner consistent with the actual search procedure, we evaluated the Lomb--Scargle periodogram of each surrogate and recorded the maximum power within a predefined symmetric search interval centered on the observed low-frequency peak, $f\in[f_{\rm peak}\pm100~{\rm Hz}]$, thereby reproducing the same local search strategy used for the observed bursts.

The empirical Monte Carlo probability was computed as
\begin{equation}
p
=
\frac{1}{N_{\rm sim}}
\sum_{j=1}^{N_{\rm sim}}
\mathbf{1}
\left[
P_{{\rm LS,sim}}^{(j)}
\ge
P_{{\rm LS,obs}}
\right],
\end{equation}
where $N_{\rm sim}$ is the number of surrogate realizations, $P_{{\rm LS,sim}}^{(j)}$ denotes the maximum Lomb--Scargle power recovered within the predefined search interval for the $j$-th realization, $P_{{\rm LS,obs}}$ is the observed Lomb--Scargle power at the low-frequency excess, and $\mathbf{1}[\cdot]$ is the indicator function. Thus, $p$ represents the fraction of simulated realizations whose strongest local excess equals or exceeds the observed one.

In principle, if periodogram powers at neighboring frequencies were statistically independent, the distribution of maxima could be approximated using generalized extreme-value statistics\cite{1928PCPS...24..180F}. In our case, however, the combination of oversampling and strong red-noise correlations breaks this assumption. We therefore estimate significance directly from the empirical distribution of simulated maxima rather than imposing an analytic extreme-value model.

Importantly, because each surrogate realization records the maximum Lomb–Scargle power recovered within the same predefined search interval, the empirical distribution naturally reflects the frequency correlations introduced by oversampling and the local search strategy adopted for the observed data. The resulting probabilities therefore provide a direct assessment of the observed excesses under the same analysis procedure.

Under the pulse-envelope-only null model, the observed low-frequency excesses are recovered with probabilities of $p=0.051$ for GRB~910711 and $p=0.001$ for GRB~200415A (Fig.~\ref{fig2}), indicating that the observed signals are not readily reproduced by pulse-envelope variability alone. However, when the previously reported kilohertz oscillatory excesses are incorporated through the QPO-informed stochastic reconstruction, the corresponding probabilities increase to $p=0.155$ and $p=0.334$, respectively. This result suggests that the low-frequency excesses become less unexpected once the reported kilohertz variability is included, consistent with a possible connection between the low- and high-frequency structures. Motivated by this correspondence, we next investigate whether these apparently separated excesses can be interpreted within a common harmonic organization.

\noindent\textbf{Statistical tests for the bispectrum.}
To quantify whether the harmonic-like structures seen in the observed bispectral maps could arise from pulse-envelope variability and counting noise, we performed Monte Carlo tests using the FRED-envelope synthetic light curves described above. The synthetic light curves were binned at $\Delta t=50~\mu$s, which is sufficient to resolve the candidate harmonic nodes up to several kilohertz while keeping the computation tractable. For each surrogate, we applied the same zero-padding factor of ten and Hamming window used for the observed light curves before computing the bispectral magnitude. Zero-padding was used only to refine the sampled frequency grid and does not improve the intrinsic Fourier resolution; the Hamming window was used to reduce sidelobe leakage from the finite $\approx$10~ms burst window. Because the identical procedure was applied to both data and simulations, the resulting null distributions include the effects of finite-duration leakage, oversampling, and red-noise contamination.

For each candidate characteristic frequency $f_0$, we evaluated the bispectral magnitude around the harmonic coupling nodes identified from the WWZ spectrograms. For a node $H(n_1,n_2)$, corresponding to target frequencies $(n_1f_0,n_2f_0)$, we defined a local bispectral statistic as
\begin{equation}
s_{n_1,n_2}(f_0)
=
\frac{
\sum_{(f_1,f_2)\in\Omega}
K(f_1,f_2)\,|\hat B(f_1,f_2)|
}{
\sum_{(f_1,f_2)\in\Omega}
K(f_1,f_2)
},
\end{equation}
where $|\hat B(f_1,f_2)|$ is the bispectral magnitude and $K(f_1,f_2)$ is the product of the Hamming-window response functions centered on $n_1f_0$, $n_2f_0$, and $(n_1+n_2)f_0$, respectively. The local region $\Omega$ was defined by retaining pixels for which the normalized kernel response exceeds one half of its peak value, thereby restricting the measurement to the local neighborhood of the target harmonic node while accounting for modest frequency offsets introduced by the short burst duration and finite spectral resolution. This construction measures the local bispectral magnitude near each nominal harmonic node while allowing for modest frequency offsets expected from the short duration and finite spectral resolution of the spike.

Rather than treating any single frequency pair as representative evidence, we combined the local node statistics into an equal-weighted joint statistic,
\begin{equation}
S(f_0)
=
\sum_{(n_1,n_2)\in\mathcal{H}}
s_{n_1,n_2}(f_0),
\end{equation}
where $\mathcal{H}$ is the set of candidate coupling nodes inferred from the observed timing structure. For GRB~910711 we used $\mathcal{H}=\{(1,1),(1,5),(5,1),(3,3),(5,6),(6,5)\}$, and for GRB~200415A we used $\mathcal{H}=\{(1,1),(2,2),(2,4),(4,2),(6,6)\}$. All candidate nodes were assigned equal weight, avoiding preferential emphasis on any single pair.

We first used this statistic to test the preferred characteristic frequency scale. For each observed burst and each surrogate realization, $S(f_0)$ was evaluated over $f_0=260$--$460$~Hz in steps of 10~Hz, and the frequency maximizing the joint statistic was recorded. Although the 10~ms spike duration limits the resolution of any isolated low-frequency feature, a harmonic-like structure provides a more stringent internal test because small changes in $f_0$ are amplified at higher multiples. In the observed data, both bursts favor a characteristic frequency close to 360~Hz despite having different sets of recovered harmonic nodes (Fig.~\ref{efig5}\textbf{A,D}). By contrast, the synthetic light curves do not show a comparable preference for 360~Hz (Fig.~\ref{efig5}\textbf{B,E}), indicating that the observed frequency scale is unlikely to be produced solely by short-window effects and variable pulse envelopes.

We then assessed the significance of the joint harmonic structure. For each surrogate realization, we recorded both the maximum joint statistic over the full $260$--$460$~Hz search interval,
\begin{equation}
S_{\rm max}
=
\max_{f_0\in[260,460]~{\rm Hz}}
S(f_0),
\end{equation}
and the joint statistic at $f_0=360$~Hz, denoted $S_{360}$. The empirical tail probability was estimated as
\begin{equation}
p
=
\frac{1}{N_{\rm sim}+1}
\left[
1+
\sum_{j=1}^{N_{\rm sim}}
\mathbf{1}
\left(
S_{\rm sim}^{(j)}
\ge
S_{\rm obs}
\right)
\right],
\end{equation}
where $S_{\rm sim}^{(j)}$ is either $S_{\rm max}$ or $S_{360}$ for the $j$-th surrogate, depending on the comparison being made. Using $S_{\rm max}$ provides the more conservative test because it incorporates the search over the full $260$--$460$~Hz interval. Under this test, the observed joint statistics correspond to probabilities of $p=1.3\times10^{-4}$ for GRB~910711 and $p=4.5\times10^{-4}$ for GRB~200415A (Fig.~\ref{efig5}\textbf{B,E}). These values indicate that the observed concentration of bispectral coupling near the harmonic nodes is rarely reproduced by the envelope-only null simulations.

Finally, we examined the individual harmonic nodes as a consistency check rather than as independent detection claims. Because the nodes are selected as parts of a candidate harmonic organization, their significances are not statistically independent and should not be interpreted as separate trials. We therefore report the marginal distributions of grouped node statistics to show how different parts of the harmonic structure contribute to the joint signal (Fig.~\ref{efig5}\textbf{C,F}). Some individual nodes are naturally less significant than others, especially at lower frequencies where red-noise variance is stronger, but the combined structure remains significant under the same null model. This behavior is expected for a harmonic-like pattern: the relevant evidence is not that every node independently exceeds a fixed threshold, but that the set of spectrogram-motivated nodes collectively favors the same characteristic frequency and exhibits enhanced bispectral coupling relative to synthetic envelope-only light curves.

Taken together, these tests separate three related but distinct statements. First, the observed harmonic-node structure favors a characteristic frequency scale near 360~Hz. Second, the joint bispectral statistic associated with this structure is unlikely to arise from the FRED-envelope null model. Third, the individual nodes provide supporting, non-independent consistency checks of the same harmonic-like organization. Within the resolution and duration permitted by the data, these results support the interpretation that the observed high-frequency excesses are organized around a common characteristic scale and are consistent with nonlinear coupling among the candidate harmonic components.

\subsection*{Energy budgets and host searches}
The $E_{\rm p, z}$--$E_{\rm iso}$ relation is a classical empirical correlation linking the rest-frame spectral hardness and isotropic energy of GRBs, commonly referred to as the ``Amati relation''\cite{2002A&A...390...81A}. Although originally established for core-collapse-driven Type~II~GRBs, related empirical relations have also been proposed for merger-driven Type~I~GRBs\cite{2020MNRAS.492.1919M} and MGFs\cite{2020ApJ...903L..32Z}, albeit with different parameter sets. For GRB~910711, the isotropic energy was calculated from the cutoff power-law model using the best-fit parameters and $1\sigma$ uncertainties integrated over the 1--10000~keV band. For GRB~200415A, we adopted the spectral parameters reported in ref.\cite{2020ApJ...899..106Y}.

As shown in Fig.~\ref{efig6}, the location of GRB~910711 in the $E_{\rm p, z}$--$E_{\rm iso}$ plane lies even further from the merger-driven GRB population than GRB~200415A when host-galaxy redshifts are treated as unknown. Furthermore, if GRB~910711 is considered within the luminosity-distance allowance of the current MGF population (including candidate events\cite{2020ApJ...903L..32Z}), we obtain an upper limit of $\approx$55~Mpc. This distance is also consistent with expectations based on BATSE's sensitivity\cite{2006ApJ...640..849N}. Motivated by this, we carried out a host-galaxy search for GRB~910711 using its reported localization\cite{1992Natur.359..217B} and the GLADE+ catalog\cite{2022MNRAS.514.1403D}, which provides an extensive compilation of nearby galaxies suitable for counterpart searches. The results, summarized in Table~\ref{etab4}, show that 17 galaxies lie within 55~Mpc of the burst position, with the nearest candidate at about $9.14\pm0.85$~Mpc. These findings demonstrate that GRB~910711 could plausibly be interpreted as an extragalactic MGF, and if so, it would represent the first such case detected by BATSE.


\clearpage

\clearpage

\clearpage


	
\clearpage
\begin{table}
\centering
\caption{\textbf{Best-fit Parameters for the spectral models of GRB~910711.} Spectral fitting was performed on the time-integrated spectrum of the burst over the 25--1800~keV energy range. All presented uncertainties correspond to the 1$\sigma$ credible interval.}
\label{tab1}
\begin{tabular}{lcccc}
\hline
Model & Parameter & Best-fit Value & PGSTAT/dof
& BIC \\
\hline
blackbody & kT (keV) & $216.27^{+43.14}_{-37.21}$ & 32.05/37 & 39.37 \\
\hline
multi-color blackbody & $kT_{\min}$ (keV)& $47.31^{+37.79}_{-42.71}$ & 31.89/35 & 46.38\\
& $kT_{\max}$ (keV) & $535.79^{+1174.21}_{-211.45}$ & & \\
& m & $-2.19^{+0.60}_{-1.03}$ & & \\
\hline
cutoff power law & $\alpha$ & $0.87^{+0.675}_{-0.687}$ & 31.72/36 & 42.71\\
& $E_p$ (keV) & $957.19^{+508.35}_{-209.02}$ & & \\
\hline
band function & $\alpha$ & $0.91^{+0.68}_{-0.70}$ & 31.07/35 & 45.70\\
& $E_p$ (keV) & $912.01^{+533.43}_{-187.57}$ & & \\
& $\beta$ & $-3.97^{+1.30}_{-1.30}$ & & \\
\hline
single power law & $\Gamma$ & $-0.87^{+0.08}_{-0.09}$ & 46.72/37 & 54.05 \\
\hline
\end{tabular}
\end{table}


\clearpage




\section*{Acknowledgments}
We thank Xing Yang for helpful comments. This work was performed on an HPC server equipped with two Intel Xeon Gold 6248 processors at Nanjing University. We acknowledge IT support from the computer lab of the School of Astronomy and Space Science at Nanjing University.
\paragraph*{Funding:}
This work was supported by the National Natural Science Foundation of China (grants 12121003, 11833003, U2038105  and 12573046 to B.-B.Z.), the National Key Research and Development Programs of China (grants 2022YFF0711404 and 2022SKA0130102 to B.-B.Z.) and the National SKA Program of China (grant 2022SKA0130100 to B.-B.Z.).
Y.L. acknowledges support from the Simons Collaboration on Extreme Electrodynamics of Compact Sources (grant number MP-SCMPS-0000147) and the Simons Investigator Grant (award number 827103).
B.-B.Z acknowledge support by the science research grants from the China Manned Space Program (grant CMS-CSST-2021-B11), the Fundamental Research Funds for the Central Universities, and the Program for Innovative Talents and Entrepreneurs in Jiangsu.
J.Y. acknowledges support from the Young Talent Cultivation Program of Zhengzhou University (Grant No. 35240910).
\paragraph*{Author contributions:}
Conceptualization: R.-C.C., O.G., B.-B.Z., and D.L.
Methodology: R.-C.C., B.-B.Z., Y.L., and S.-N.W.
Resources: R.-C.C., B.-B.Z., and J.Y.
Investigation: R.-C.C., O.G., B.-B.Z., Y.L., D.L., Y.-H.I.Y., and S.-N.W.
Data curation: R.-C.C. and B.-B.Z.
Software: R.-C.C. and B.-B.Z.
Formal analysis: R.-C.C., B.-B.Z., and J.Y.
Validation: R.-C.C., O.G., and B.-B.Z.
Visualization: R.-C.C., B.-B.Z., and J.Y.
Writing -- original draft: R.-C.C., O.G., B.-B.Z., and J.Y.
Writing -- review \& editing: R.-C.C., O.G., B.-B.Z., Y.L., D.L., and Y.-H.I.Y.
Supervision: R.-C.C., O.G., and B.-B.Z.
Project administration: R.-C.C. and B.-B.Z.
Funding acquisition: R.-C.C., B.-B.Z., and Y.L.
\paragraph*{Competing interests:}
There are no competing interests to declare.
\paragraph*{Data, Code, and Materials Availability:}
All data and code needed to evaluate and reproduce the results in the paper are present in the paper and/or the Supplementary Materials. This study did not generate new materials.
All data used in this study are publicly available through mission archives or published repositories.
BATSE TTE data are available at \url{https://heasarc.gsfc.nasa.gov/FTP/compton/data/batse/ascii_data/batse_tte/}.
BATSE STTE data are available at \url{https://heasarc.gsfc.nasa.gov/FTP/compton/data/batse/trigger/}.
Swift-BAT data can be accessed via HEASARC (\url{https://heasarc.gsfc.nasa.gov/W3Browse/swift/swiftgrb}) and the GUANO repository (\url{https://www.swift.psu.edu/guano}).
ASIM data are available from Zenodo (\url{https://doi.org/10.5281/zenodo.5242975}).
All analysis codes (mostly in \texttt{Python}), implementation details, and parameters are fully described in the paper and the Supplementary materials.


\section*{Supplementary materials}
Supplementary Text\\
Figs. S1 to S6\\
Tables S1 to S4\\
References \textit{(79-\arabic{enumiv})} 


\clearpage
\pagestyle{empty}
\thispagestyle{empty}


\renewcommand{\thefigure}{S\arabic{figure}}
\renewcommand{\thetable}{S\arabic{table}}
\renewcommand{\theequation}{S\arabic{equation}}
\setcounter{figure}{0}
\setcounter{table}{0}
\setcounter{equation}{0}
\setcounter{page}{1} 


\begin{center}
\section*{Supplementary Materials for\\ \scititle}


Run-Chao Chen$^{\ast}$,
Ore Gottlieb$^{\ast}$,
Bin-Bin Zhang$^{\ast}$,
Yuri Levin,
Dong Lai,
Jun Yang,
Yi-Han Iris Yin,
Suo-Ning Wang \\
\small$^\ast$Corresponding authors: R.-C. C. (\url{chrczxx@smail.nju.edu.cn});
O. G. (\url{ogot@mit.edu});
B.-B. Z. (\url{bbzhang@nju.edu.cn}).
\end{center}

\subsubsection*{This PDF file includes:}
Supplementary Text\\
Figures S1 to S6\\
Tables S1 to S4\\


\newpage


\section*{Supplementary Text}

\subsection*{Exploratory Discussion on Possible Physical Interpretations of the $\approx$360~Hz Characteristic Scale}

The apparent preference for a characteristic scale near 360~Hz in both GRB~910711 and GRB~200415A remains puzzling, as the magnetospheres during MGFs are expected to be highly disturbed and are unlikely to support nearly identical hydromagnetic modes, although the excitation nodes differ between the two events, potentially owing to signal-to-noise limitations or intrinsic physical differences. If the oscillation clocks are instead associated with the vibrational modes of the neutron stars themselves, this raises interesting questions about what these modes are, how they are excited, and whether the initial spikes of extragalactic MGFs can be used for magnetar seismology. Observational evidence for high-frequency QPOs extending to several hundred hertz has been reported in the pulsating tails of the giant flare from the Galactic magnetar SGR~1806$-$20\cite{2006ApJ...637L.117W,2006ApJ...653..593S}, where a $\approx$625~Hz QPO was attributed to a transient $n=1$ crustal shear mode\cite{2005ApJ...634L.153P,2014ApJ...793..129H}. If the nuclear symmetry energy varies strongly with density near saturation\cite{2009PhRvL.103r1101S}, the expected $n=1$ frequency would be lower (e.g., by a factor of $\approx$0.5--0.6), yielding a value consistent with the 360~Hz oscillation observed here. However, it should be noted that the frequencies of crustal shear modes can vary substantially with the uncertainties in the equation of state, magnetic field configuration, and stellar structure, making it difficult to explain the nearly identical 360~Hz frequencies in the two MGFs as anything other than a coincidence.

Alternatively, another exciting (albeit speculative) possibility is that the 360 Hz QPOs observed in both MGFs correspond to a new type of oscillation modes anchored deep inside the neutron star, with the mode frequency depending mainly on the fundamental physics of nuclear matter and weakly on the specifics of the neutron star (e.g., mass). One such mode is the interface mode associated with a hadronic matter to quark matter phase transition. The possibility of deconfined quark matter inside the neutron star core has been discussed for many decades\cite{2005ApJ...629..969A}.  The frequency of such an interface mode depends on the transition density $\rho_{\rm tr}$ and the density jump $\Delta\rho$ across the phase transition\cite{1987MNRAS.227..265F}, both of which depend only on nuclear physics. For a reasonable $\Delta\rho/\rho_{\rm tr}$, the mode frequency lies in the range of 100--1000~Hz\cite{2024ApJ...964...31M}, and thus could account for the 360 Hz frequency observed in both magnetars.



\clearpage
\begin{figure}
\centering
\includegraphics[width=0.75\textwidth]{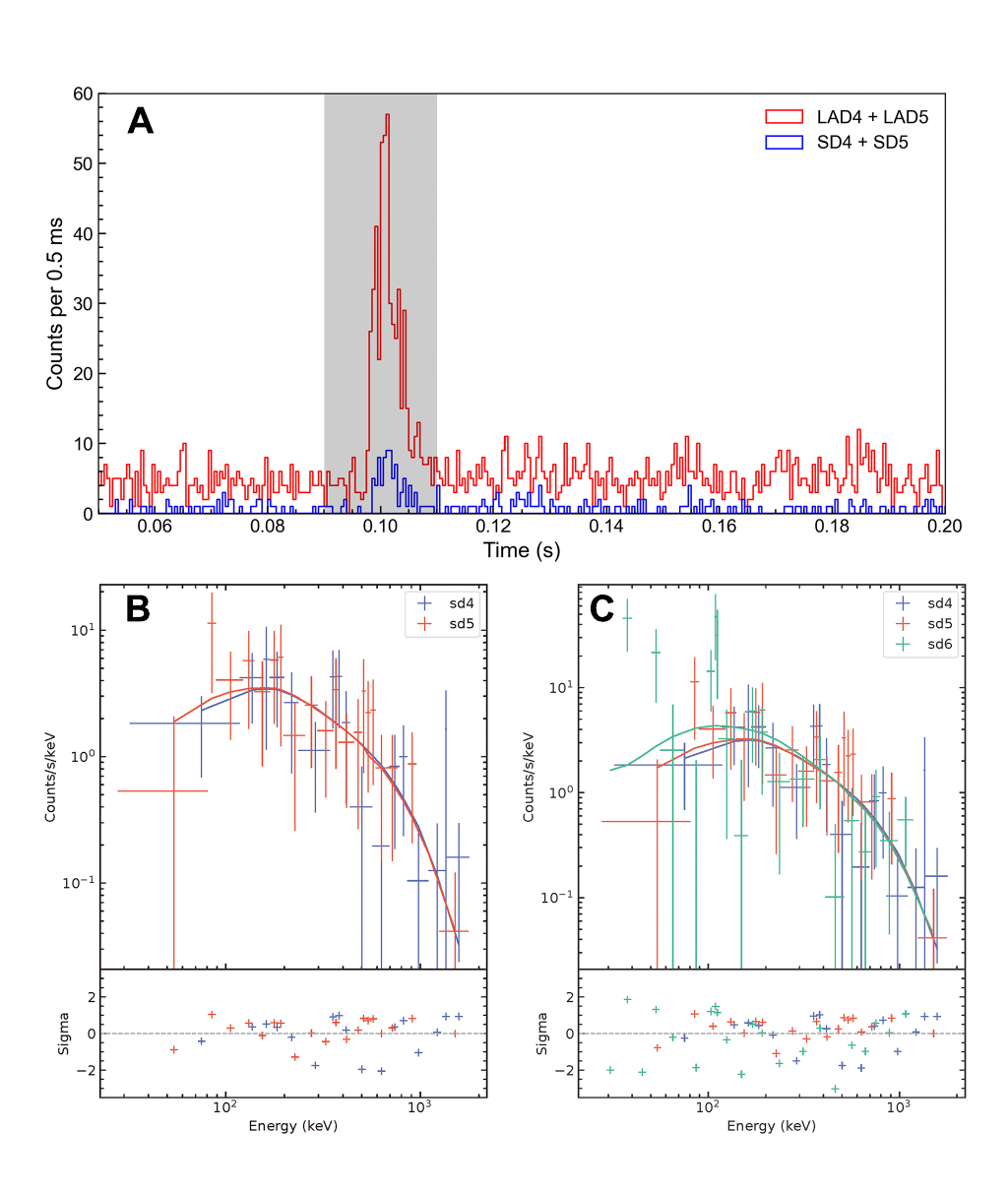}
\caption{\textbf{Time-integrated spectral fitting of GRB~910711 using spectroscopy detector data.}
(\textbf{A}) Light curves recorded by both the LADs and SDs across their full energy ranges are shown. The time interval selected for spectral analysis is marked with the gray band. 
(\textbf{B}) Photon count spectra fitted with the best-fitting blackbody model, excluding data from SD6. 
(\textbf{C}) Photon count spectra fitted with the best-fitting blackbody model, including data from SD6. 
Although SD6 observed the burst at a reasonable source angle of approximately 55$^{\circ}$, its small Earth angle of about 78$^{\circ}$ resulted in significant photon scattering. This contamination alters the recorded energy spectrum and causes the best-fit model to deviate from thermal to a simple power-law shape.}
\label{efig1}
\end{figure}

\clearpage
\begin{figure}
\centering
\includegraphics[width=\textwidth]{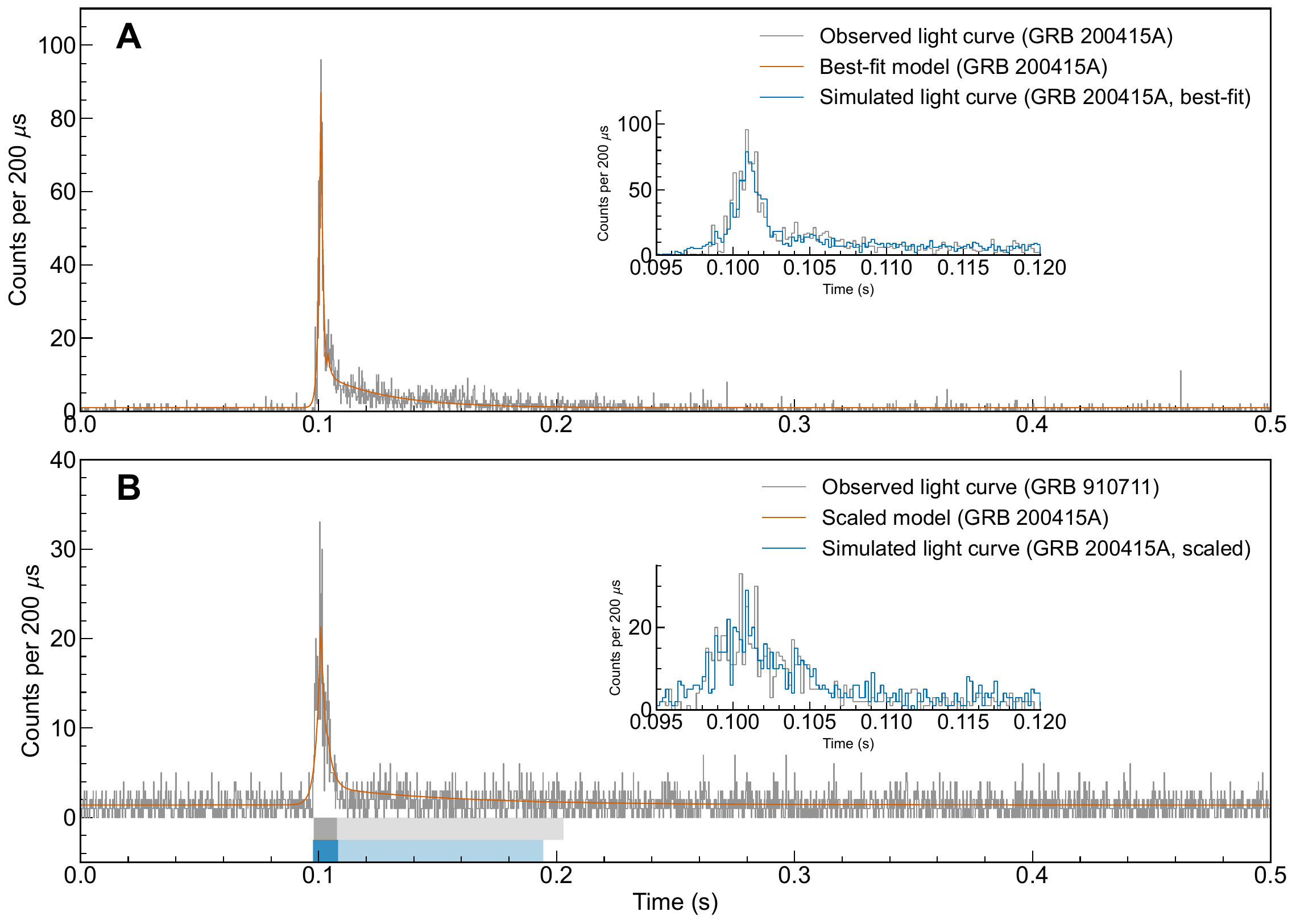}
\caption{\textbf{Reconstruction of the GRB~910711 light curve using a template derived from GRB~200415A.}
Gray curves show the observed light curves in the shared 60--325~keV energy band for GRB~200415A (Swift-BAT, panel~A) and GRB~910711 (BATSE-LADs, panel~B). Insets highlight the initial spikes; blue lines show the corresponding simulated light curves. 
(\textbf{A}) The orange line indicates the best-fit model for GRB~200415A. 
(\textbf{B}) The orange line shows a scaled version of the model in panel~A, assuming a 40\% increase in detector sensitivity and placing GRB~200415A at an effective distance of $\approx$8.75~Mpc. Shaded regions indicate the time intervals of the initial spike and the decaying tail, as determined by Bayesian blocks, in both the observed (gray) and simulated (blue) light curves.}
\label{efig2}
\end{figure}

\clearpage
\begin{figure}
\centering
\includegraphics[width=0.5\textwidth]{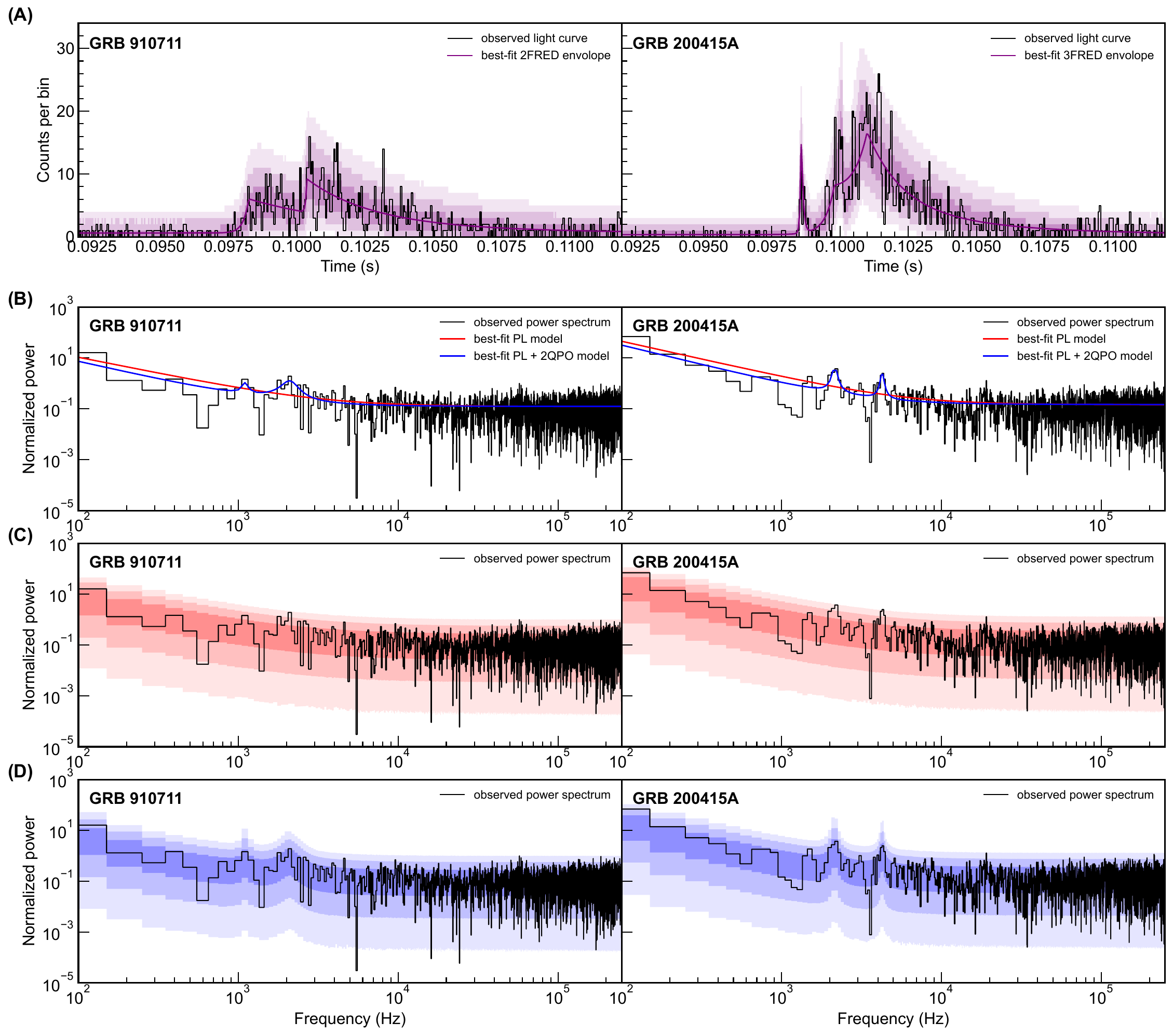}
\caption{\textbf{Light-curve and power-spectrum reconstruction for GRB~910711 and GRB~200415A.}
\textbf{(A)} FRED-envelope modeling of the initial spikes. The observed light curves in the interval [92, 112]~ms are shown together with the preferred multi-FRED envelope models (purple curves; two FREDs for GRB~910711 and three FREDs for GRB~200415A). Shaded regions indicate the central 1, 2, 3$\sigma$ regions of synthetic light curves generated from the posterior distributions of the fitted envelope models after Poisson sampling, demonstrating that the short-timescale burst morphology is reproduced across the posterior parameter space.
\textbf{(B)} Observed power spectra (black) compared with the best-fit continuum-only (PL+C; red) and continuum-plus-oscillation (PL+C+2QPO; blue) models derived from 2-$\mu$s binned light curves. The additional Lorentzian components recover the previously reported kilohertz excesses and provide an improved description of the localized high-frequency structure.
\textbf{(C)} Power-spectrum reconstruction using synthetic light curves generated from the PL+C model. Colored bands denote the central 1, 2, 3$\sigma$ ranges of reconstructed power spectra from $10^5$ T--K synthesis followed by Poisson sampling. While the continuum-only model broadly reproduces the red-noise behavior, it tends to overestimate low-frequency variance and under-reproduce the observed localized high-frequency excesses.
\textbf{(D)} Same as \textbf{(C)}, but for the PL+C+2QPO model. Incorporating the previously reported kilohertz oscillatory excesses yields synthetic realizations that better reproduce both the broad continuum and the observed localized power excesses. These reconstructions motivate the use of both FRED-envelope and QPO-informed stochastic null models in the subsequent timing analyses.}
\label{efig3}
\end{figure}

\clearpage
\begin{figure}
\centering
\includegraphics[width=0.75\textwidth]{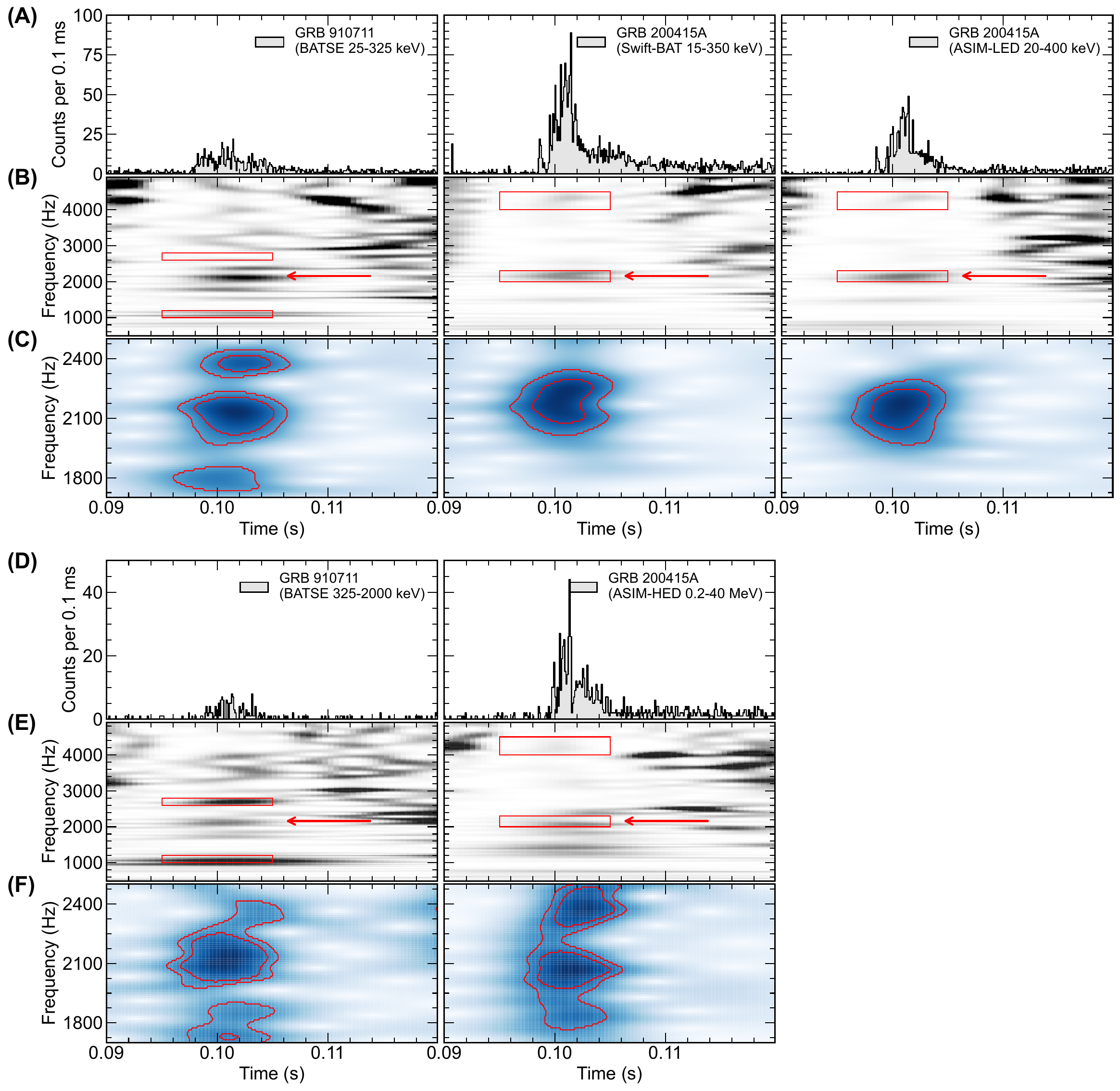}
\caption{\textbf{Spectrograms of GRB~910711 and GRB~200415A.}
Light curves used for spectrogram analysis are binned at 100~$\mu$s resolution, with a frequency resolution of 10~Hz. The exponential decay constant is set as $c_w = \frac{1}{(2000~\text{Hz} \times 0.03~\text{s})^2}$. 
(\textbf{A}) Low-energy light curves of GRB~910711 (BATSE-LADs, energy channels 1--3) and GRB~200415A (Swift-BAT and ASIM-LED). 
(\textbf{B}) WWZ spectrograms derived from the low-energy light curves, computed with 1-ms time resolution. Darker regions indicate higher statistical significance of periodic signals. Red boxes denote the previously reported kHz QPOs\cite{2021Natur.600..621C,2023Natur.613..253C}; red arrows mark the shared $\approx$2160~Hz QPO seen in both bursts. 
(\textbf{C}) WWA spectrograms for the same light curves, computed with 0.2-ms time resolution. Darker regions indicate stronger modulation power. Red contours enclose the top 25\% and 50\% of the total WWA energy, highlighting the most energetically significant QPO components. 
(\textbf{D}) High-energy light curves of GRB~910711 (BATSE-LADs, energy channel 4) and GRB~200415A (ASIM-HED). 
(\textbf{E}) Same as panel~B, but for the high-energy light curves. 
(\textbf{F}) Same as panel~C, but for the high-energy light curves.}
\label{efig4}
\end{figure}

\clearpage
\begin{figure}
\centering
\includegraphics[width=0.48\textwidth]{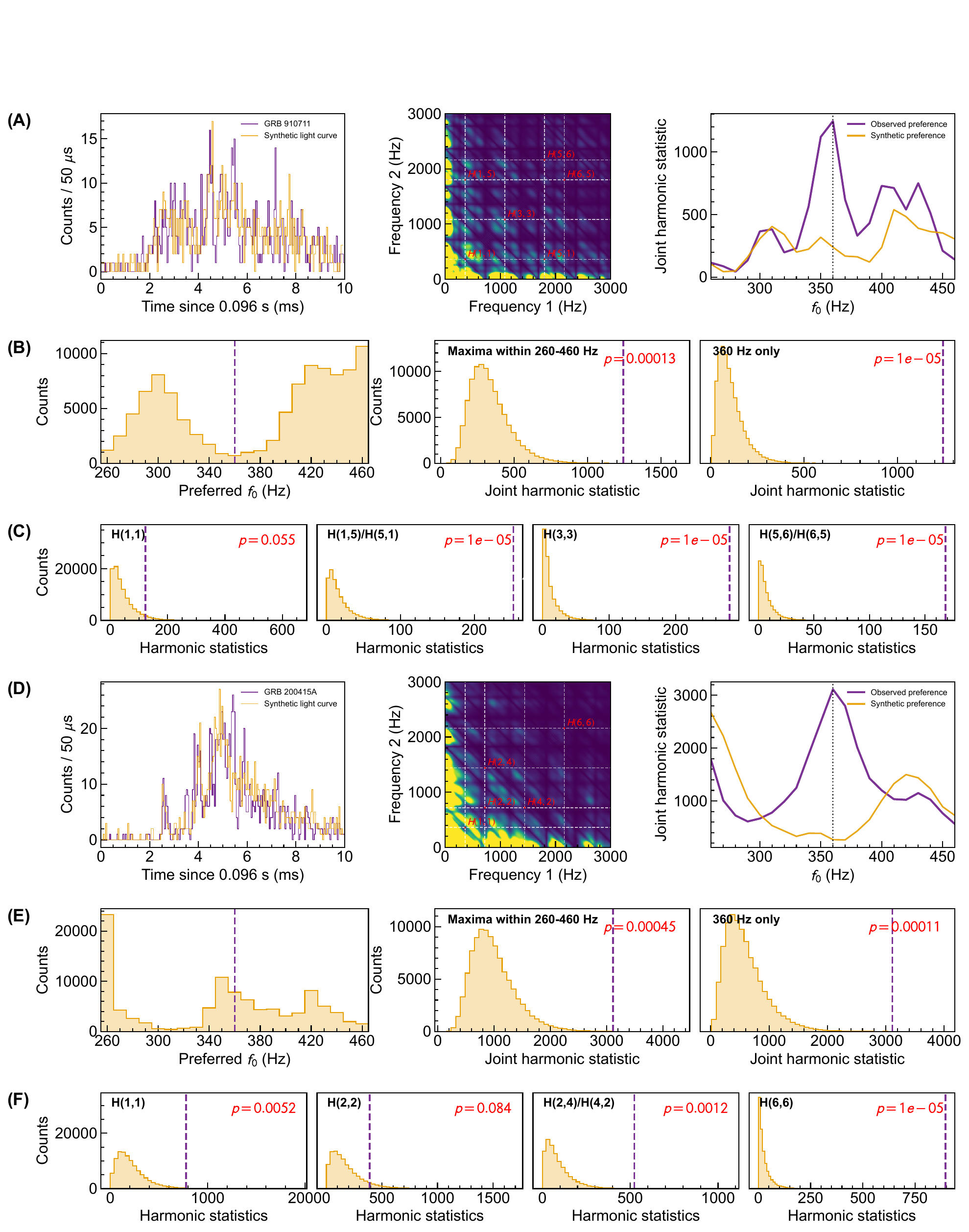}
\caption{\textbf{Monte Carlo tests of the bispectral harmonic-node structure.}
\textbf{(A)} Example comparison between the observed light curve and one synthetic FRED-envelope realization, together with the corresponding bispectral map and the joint harmonic statistic $S(f_0)$. Red contours mark the local regions around the candidate harmonic coupling nodes used to evaluate the bispectral statistics. The observed data show a preference near $f_0=360$~Hz, while the example synthetic realization does not reproduce the same harmonic-node pattern.
\textbf{(B)} Monte Carlo distributions from $10^5$ FRED-envelope synthetic light curves. The left panel shows the distribution of the preferred characteristic frequency, defined as the $f_0$ that maximizes the joint harmonic statistic over 260--460~Hz. The middle panel shows the distribution of the maximum joint statistic over this interval, and the right panel shows the distribution of the joint statistic evaluated specifically at $f_0=360$~Hz. Purple dashed lines mark the observed values.
\textbf{(C)} Marginal distributions of grouped harmonic-node statistics at $f_0=360$~Hz. These panels show how different parts of the candidate harmonic structure contribute to the observed joint statistic.
\textbf{(D--F)} Same as \textbf{(A--C)}, but for GRB~200415A. Despite differences in the specific harmonic nodes recovered in the two bursts, both observed events favor a characteristic frequency close to 360~Hz and show enhanced joint harmonic statistics relative to the FRED-envelope null simulations.}
\label{efig5}
\end{figure}

\clearpage
\begin{figure}
\centering
\includegraphics[width=0.85\textwidth]{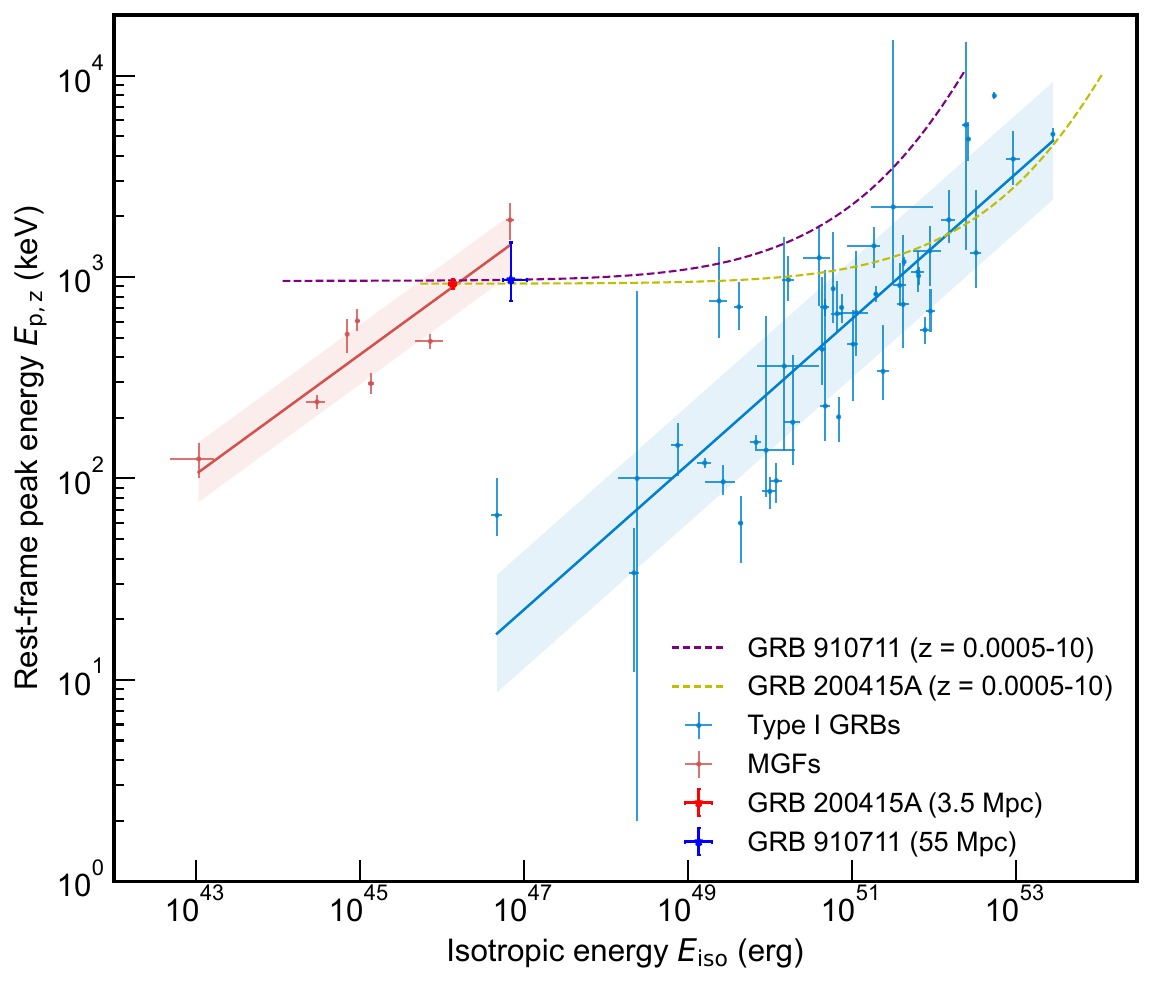}
\caption{\textbf{Comparison of GRB~910711 with merger-driven GRBs and MGF-associated bursts in the $E_{\rm iso}$--$E_{\rm p,z}$ plane.}
Light-blue points show merger-driven (Type~I) GRBs with 1$\sigma$ uncertainties, taken from ref.\cite{2020MNRAS.492.1919M}. Light-red points represent MGFs and candidate short GRBs proposed as extragalactic MGFs associated with nearby host galaxies, compiled from refs.\cite{2020ApJ...903L..32Z,2020ApJ...899..106Y,2024ApJ...963L..10Y}. Solid lines denote the best-fitting empirical correlations for the two populations, with shaded regions indicating the corresponding 1$\sigma$ confidence intervals. 
The red error bar highlights GRB~200415A, located in the nearby galaxy NGC~253 at a luminosity distance of 3.5~Mpc. 
For GRB~910711 and GRB~200415A, if assumed as short GRBs which lack precise spectroscopic redshifts due to the absence of optical afterglows, possible locations are shown as purple and yellow dashed tracks spanning $z=0.0005$--$10$. 
A representative point for GRB~910711, highlighted in blue, corresponds to a maximum luminosity distance of 55~Mpc, consistent with the distribution of MGFs.}
\label{efig6}
\end{figure}

\clearpage

\begin{table}
\centering
\caption{\textbf{Best-fit Parameters for the model fitting of the overall light curves of GRB~200415A and GRB~910711.} Parameters marked with a logarithmic symbol indicate that a base-10 logarithmic prior was used during the fitting process. All parameter units are expressed in a linear (non-logarithmic) scale. The $t_{\text{peak}}$ values are presented in milliseconds for clarity in conveying their associated uncertainties. All presented uncertainties correspond to the 1$\sigma$ credible interval.}
\label{etab1}
\begin{tabular}{lccc}
\hline
Parameter & Prior range & Best-fit Value & Unit \\
\hline
$\log_{10} A_1$ & [0, 3] & $1.950^{+0.003}_{-0.003}$ & counts/200$\mu$s \\
$t_{\text{peak}1}$ & [90, 102] & $100.934^{+0.008}_{-0.008}$ & ms \\
$\log_{10} \tau_{\text{rise}1}$ & [-4, -2] & $-3.029^{+0.005}_{-0.005}$ & s \\
$\log_{10} \tau_{\text{decay}1}$ & [-4, -2] & $-2.948^{+0.007}_{-0.007}$ & s \\
$\log_{10} A_2$ & [0, 2] & $0.927^{+0.010}_{-0.010}$ & counts/200$\mu$s \\
$t_{\text{peak}2}$ & [102, 150] & $103.798^{+0.053}_{-0.045}$ & ms \\
$\log_{10} \tau_{\text{rise}2}$ & [-4, -1] & $-3.597^{+0.075}_{-0.091}$ & s \\
$\log_{10} \tau_{\text{decay}2}$ & [-3, 0] & $-1.580^{+0.016}_{-0.016}$ & s \\
$C$ & [0, 10] & $0.991^{+0.013}_{-0.013}$ & counts/200$\mu$s \\
\hline
$a$ & [0.01, 1] & $0.227^{+0.005}_{-0.004}$ & - \\
$b$ & [0.1, 10] & $1.419^{+0.015}_{-0.015}$ & - \\
$c$ & [0.01, 1] & $0.480^{+0.013}_{-0.013}$ & - \\
\hline
\end{tabular}
\end{table}

\clearpage
\begin{table}
\scriptsize
\centering
\caption{\textbf{Best-fit parameters for the FRED envelope modeling of the spike light curves of GRB~910711 and GRB~200415A.}
All presented uncertainties correspond to the 1$\sigma$ credible interval.}
\label{etab2}
\begin{tabular}{lccc}
\hline
Parameter & Prior range & Best-fit Value & Unit \\
\hline
\multicolumn{4}{l}{\textbf{GRB~910711: 2-FRED model}} \\
$A_1$ & [1, 15] & $5.952^{+1.177}_{-0.763}$ & counts/50$\mu$s \\
$t_{\rm peak1}$ & [96, 100] & $98.463^{+0.286}_{-0.252}$ & ms \\
$\log_{10}\tau_{\rm rise1}$ & [-4.7, -2.0] & $-3.508^{+0.189}_{-0.290}$ & s \\
$\log_{10}\tau_{\rm decay1}$ & [-4.7, -2.0] & $-2.366^{+0.064}_{-0.126}$ & s \\
$A_2$ & [1, 15] & $5.835^{+4.566}_{-1.118}$ & counts/50$\mu$s \\
$t_{\rm peak2}$ & [100, 106] & $100.393^{+0.147}_{-0.059}$ & ms \\
$\log_{10}\tau_{\rm rise2}$ & [-4.7, -2.0] & $-4.130^{+0.354}_{-0.348}$ & s \\
$\log_{10}\tau_{\rm decay2}$ & [-4.7, -2.0] & $-2.728^{+0.171}_{-1.659}$ & s \\
$C$ & [0, 26] & $0.580^{+0.077}_{-0.072}$ & counts/50$\mu$s \\
\hline
\multicolumn{4}{l}{\textbf{GRB~200415A: 3-FRED model}} \\
$A_1$ & [1, 25] & $11.737^{+3.848}_{-3.077}$ & counts/50$\mu$s \\
$t_{\rm peak1}$ & [98, 99] & $98.555^{+0.022}_{-0.016}$ & ms \\
$\log_{10}\tau_{\rm rise1}$ & [-4.7, -2.0] & $-4.484^{+0.166}_{-0.129}$ & s \\
$\log_{10}\tau_{\rm decay1}$ & [-4.7, -2.0] & $-3.941^{+0.133}_{-0.125}$ & s \\
$A_2$ & [1, 25] & $12.578^{+3.408}_{-2.941}$ & counts/50$\mu$s \\
$t_{\rm peak2}$ & [99.0, 100.5] & $100.031^{+0.045}_{-0.177}$ & ms \\
$\log_{10}\tau_{\rm rise2}$ & [-4.7, -2.0] & $-3.587^{+0.101}_{-0.174}$ & s \\
$\log_{10}\tau_{\rm decay2}$ & [-4.7, -2.0] & $-4.135^{+1.026}_{-0.394}$ & s \\
$A_3$ & [1, 25] & $16.584^{+1.407}_{-2.693}$ & counts/50$\mu$s \\
$t_{\rm peak3}$ & [100.5, 106.0] & $100.791^{+0.090}_{-0.078}$ & ms \\
$\log_{10}\tau_{\rm rise3}$ & [-4.7, -2.0] & $-3.315^{+0.107}_{-0.134}$ & s \\
$\log_{10}\tau_{\rm decay3}$ & [-4.7, -2.0] & $-2.678^{+0.060}_{-0.053}$ & s \\
$C$ & [0, 36] & $0.368^{+0.095}_{-0.086}$ & counts/50$\mu$s \\
\hline
\end{tabular}
\end{table}

\clearpage
\begin{table}
\centering
\caption{\textbf{Power-spectrum fitting results for GRB~910711 and GRB~200415A.} The table reports posterior medians and 1$\sigma$ credible intervals for the power-law continuum model (PL+C) and the power-law continuum plus two Lorentzian QPO model (PL+C+2QPO), together with the maximum Whittle log-likelihood, $\log\mathcal{L}$, for each model.}
\label{etab3}
\begin{tabular}{llcccc}
\hline
Burst & Model & $\ln\mathcal{L}$ & Parameter & Prior & Value \\
\hline
\textbf{GRB~910711} & PL+C & 2628.68 & $\ln A$ & unbounded & $8.20^{+0.57}_{-0.55}$ \\
 &  &  & $\alpha$ & $[0,\,5]$ & $1.27^{+0.07}_{-0.07}$ \\
 &  &  & $\ln C$ & unbounded & $-2.10^{+0.01}_{-0.01}$ \\
\cline{2-6}
 & PL+C+2QPO & 2635.40 & $\ln A$ & unbounded & $8.55^{+2.72}_{-1.08}$ \\
 &  &  & $\alpha$ & $[0,\,5]$ & $1.43^{+0.53}_{-0.17}$ \\
 &  &  & $\ln C$ & unbounded & $-2.09^{+0.02}_{-0.02}$ \\
 &  &  & $\ln A_1$ & $[-5.2,\,4.8]$ & $-0.41^{+1.06}_{-1.69}$ \\
 &  &  & $f_1$ (Hz) & $[1071,\,1154]$ & $1108.14^{+26.99}_{-24.70}$ \\
 &  &  & $\ln w_1$ & $[3.9,\,5.7]$ & $4.55^{+0.63}_{-0.44}$ \\
 &  &  & $\ln A_2$ & $[-4.5,\,5.5]$ & $0.05^{+0.45}_{-0.42}$ \\
 &  &  & $f_2$ (Hz) & $[1500,\,3000]$ & $2091.03^{+108.33}_{-87.97}$ \\
 &  &  & $\ln w_2$ & $[3.9,\,6.6]$ & $6.24^{+0.28}_{-0.53}$ \\
\hline
\textbf{GRB~200415A} & PL+C & 2132.00 & $\ln A$ & unbounded & $10.30^{+0.47}_{-0.47}$ \\
 &  &  & $\alpha$ & $[0,\,5]$ & $1.41^{+0.06}_{-0.06}$ \\
 &  &  & $\ln C$ & unbounded & $-1.93^{+0.01}_{-0.01}$ \\
\cline{2-6}
 & PL+C+2QPO & 2148.97 & $\ln A$ & unbounded & $10.51^{+0.94}_{-0.83}$ \\
 &  &  & $\alpha$ & $[0,\,5]$ & $1.53^{+0.14}_{-0.12}$ \\
 &  &  & $\ln C$ & unbounded & $-1.91^{+0.02}_{-0.02}$ \\
 &  &  & $\ln A_1$ & $[-3.9,\,6.1]$ & $1.44^{+0.91}_{-0.73}$ \\
 &  &  & $f_1$ (Hz) & $[2006,\,2306]$ & $2151.05^{+40.26}_{-42.72}$ \\
 &  &  & $\ln w_1$ & $[3.9,\,6.9]$ & $4.94^{+0.59}_{-0.60}$ \\
 &  &  & $\ln A_2$ & $[-4.2,\,5.8]$ & $0.89^{+0.95}_{-0.74}$ \\
 &  &  & $f_2$ (Hz) & $[3933,\,4579]$ & $4258.34^{+50.06}_{-49.59}$ \\
 &  &  & $\ln w_2$ & $[3.9,\,7.6]$ & $5.30^{+0.71}_{-0.71}$ \\
\hline
\end{tabular}
\end{table}

\clearpage
\begin{table}
\centering
\caption{\textbf{Possible host galaxies for GRB~910711.} 
The first column lists the GLADE identifiers (IDs) for each galaxy\cite{2022MNRAS.514.1403D} located within the localization uncertainty region of GRB~910711\cite{1992Natur.359..217B}, followed by the J2000 equatorial coordinates, angular separation from the GRB~910711 localization center and luminosity distance ($D_{\rm L}$).}
\label{etab4}
\begin{tabular}{lcccc}
\hline
ID & RA & Dec & Angular & $D_{\rm L}$ \\
& & & Separation ($^\circ$) & (Mpc) \\
\hline
14423097 & 214.63 & -14.40 & 3.66 & $9.14 \pm 0.85$ \\
14478844 & 215.03 & -16.93 & 2.37 & $16.63 \pm 0.52$ \\
1162365 & 213.41 & -17.98 & 4.08 & $20.49 \pm 0.8$ \\
14478851 & 214.97 & -16.91 & 2.42 & $22.94 \pm 0.59$ \\
14378354 & 213.36 & -15.92 & 4.07 & $29.17 \pm 1.0$ \\
1326604 & 214.34 & -16.63 & 3.03 & $30.55 \pm 1.04$ \\
1170296 & 213.35 & -16.53 & 3.98 & $33.03 \pm 0.89$ \\
2467037 & 212.45 & -14.79 & 5.26 & $33.57 \pm 0.85$ \\
2467038 & 212.42 & -14.77 & 5.29 & $33.93 \pm 0.85$ \\
14424082 & 214.81 & -19.83 & 3.96 & $34.55 \pm 0.74$ \\
14377564 & 213.48 & -16.74 & 3.85 & $39.16 \pm 0.88$ \\
14318017 & 212.99 & -20.31 & 5.53 & $39.25 \pm 0.83$ \\
14505912 & 215.22 & -15.80 & 2.40 & $40.62 \pm 1.01$ \\
14390531 & 213.82 & -17.93 & 3.69 & $42.10 \pm 0.92$ \\
14280431 & 212.31 & -16.22 & 5.01 & $42.76 \pm 0.95$ \\
14390597 & 214.20 & -17.85 & 3.32 & $51.25 \pm 0.77$ \\
14399148 & 214.48 & -13.94 & 4.08 & $53.86 \pm 0.84$ \\
\hline
\end{tabular}
\end{table}

\end{document}